\documentclass[a4paper,12pt]{article}

\usepackage{graphicx}
\usepackage{hyperref}

\usepackage{a4wide}

\usepackage{amsmath,amssymb}
\usepackage{url}
\usepackage{cite}

\makeatletter

\renewcommand\section{\@startsection {section}{1}{\z@}%
                  {-3.0ex \@plus +1ex \@minus +.2ex}%
                  {2.3ex \@plus.2ex}%
                  {\normalfont\large\bfseries}}
\renewcommand\subsection{\@startsection{subsection}{2}{\z@}%
                   {-2.5ex\@plus +1ex \@minus +.2ex}%
                   {1.5ex \@plus .2ex}%
                   {\normalfont\normalsize\bfseries}}

\renewcommand\paragraph{\@startsection{paragraph}{4}{\z@}{-2.00ex plus
 +1ex minus +.2ex}{1.5ex plus .2ex}{\it\normalsize}}

\makeatother

\allowdisplaybreaks

\newcommand{\ii}{\mathrm{i}}
\newcommand{\rh}{\scriptscriptstyle\mathrm{H}}
\newcommand{\ra}{\scriptscriptstyle{\mathrm{A}}}

\def\bra#1{\mathinner{\langle{#1}|}}
\def\ket#1{\mathinner{|{#1}\rangle}}
\def\braket#1{\mathinner{\langle{#1}\rangle}}
\def\sbraket#1{\mathinner{\lbrack{#1}\rbrack}}
\def\sbra#1{\mathinner{\lbrack{#1}|}}
\def\sket#1{\mathinner{|{#1}\rbrack}}

\makeatletter
\newcommand{\fmslash}[2][0mu]{%
 \mathchoice
  {\fmsl@sh\displaystyle{#1}{#2}}%
  {\fmsl@sh\textstyle{#1}{#2}}%
  {\fmsl@sh\scriptstyle{#1}{#2}}%
  {\fmsl@sh\scriptscriptstyle{#1}{#2}}}
\newcommand{\fmsl@sh}[3]{%
 \m@th\ooalign{$\hfil#1\mkern#2/\hfil$\crcr$#1#3$}}
\makeatother

\allowdisplaybreaks
\numberwithin{equation}{section}

\begin{document}
\thispagestyle{empty}
\setcounter{page}{0}

\begin{flushright}
{\small
TTK-19-43 \\
October 29, 2019}
\end{flushright}

\vspace{\baselineskip}

\begin{center}
\textbf{\large On-shell constructibility of Born amplitudes\\ in spontaneously
 broken gauge theories}\\
\vspace{3\baselineskip}
{\sc 
Robert Franken$^a$, Christian Schwinn$^b$
}\\
\vspace{0.7cm}
{\sl
${}^a$ Institut f\"ur Theoretische~Physik und Astrophysik,
 Universit\"at~W\"urzburg, D-97074~W\"urzburg, Germany
\\ \vspace{0.3cm} 
${}^b$Institut f\"ur Theoretische Teilchenphysik und  
Kosmologie, 
RWTH Aachen University, D--52056 Aachen, Germany
}

\vspace*{1.2cm}
\textbf{Abstract}\\ 

\vspace{1\baselineskip}
\parbox{0.9\textwidth}{ We perform a comprehensive study of on-shell recursion
 relations for Born amplitudes in spontaneously broken gauge theories and
 identify the minimal shifts required to construct amplitudes with a given
 particle content and spin quantum numbers. We show that two-line or
 three-line shifts are sufficient to construct all amplitudes with five or
 more particles, apart from amplitudes involving longitudinal vector bosons
 or scalars, which may require at most five-line shifts. As an application, we
 revisit selection rules for multi-boson amplitudes using on-shell recursion
 and little-group transformations. }
\end{center}
\newpage

\section{Introduction}

The development of on-shell recursion relations of Born amplitudes in gauge
theories by Britto, Cachazo, Feng, and
Witten~(BCFW)~\cite{Britto:2004ap,Britto:2005fq} has motivated an approach to
Quantum Field Theory that aims at the construction of amplitudes solely in
terms of on-shell building blocks. For massless theories, space-time
symmetries and factorization properties are sufficient to fix the structure of
three- and four-point amplitudes and to establish the uniqueness of
non-abelian gauge theories~\cite{Benincasa:2007xk}. The three-point vertices
serve as input to the BCFW recursion relations, which are based on a
continuation of amplitudes into the complex plane by a complex shift of two
external momenta. This provides a purely on-shell construction of Born
amplitudes in unbroken gauge theories. The on-shell constructibility of gauge
theories with general matter content or of Effective Field Theories was
investigated using generalizations of the BCFW construction to shifts of more
than two legs~\cite{Cohen:2010mi,Cheung:2015cba,Cheung:2015ota}.

Concerning theories with massive particles, symmetry constraints on
three-point amplitudes were obtained using
supersymmetry~\cite{Craig:2011ws,Boels:2011zz} and little-group
transformations~\cite{Chen:2011ve,Conde:2016vxs}. More recently, this analysis
was simplified using a manifestly little-group covariant notation for massive
amplitudes~\cite{Arkani-Hamed:2017jhn}. The uniqueness of four-point
amplitudes can be argued to arise from factorization properties and
consistency with the high-energy limit~\cite{Arkani-Hamed:2017jhn}. This line
of argument is expected to reproduce the classic results on the uniqueness of
spontaneously broken gauge
theories~(SBGTs)~\cite{Cornwall:1973tb,Cornwall:1974km,LlewellynSmith:1973yud}
within an on-shell approach. All three-point vertices and amplitudes for some
three- and four-body decays in the electroweak Standard Model~(SM) were
constructed using this
formalism~\cite{Christensen:2018zcq,Christensen:2019mch}, while all
three-point vertices including contributions from higher-dimensional operators
were obtained in~\cite{Durieux:2019eor}.

In the current paper we investigate the on-shell constructibility of
higher-point Born amplitudes in SBGTs. After initial explorations of BCFW
recursion for amplitudes with massive
particles~\cite{Badger:2005zh,Badger:2005jv,Forde:2005ue,Ferrario:2006np,Ozeren:2006ft},
it was shown that amplitudes in QCD with massive quarks can be constructed
using at most three-line shifts~\cite{Schwinn:2007ee} whereas the on-shell
constructibility of amplitudes in massive, power-counting renormalizable
theories using all-line shifts was proven in~\cite{Cohen:2010mi}. The
constructibility of massless amplitudes with the matter content of SBGTs using
at most five-line shifts was demonstrated in~\cite{Cheung:2015cba}. We extend
this analysis to the broken phase and show that all amplitudes involving at
least two transverse vector bosons can be constructed using three-line shifts,
while amplitudes with longitudinal vector bosons or scalars may require
four- or five-line shifts. This requires the analysis of the behaviour of
amplitudes for large values of the parameter $z$, which parameterises the
complex continuation of the amplitude. Because of the analogy of the
large-$z$ limit with a high-energy limit~\cite{ArkaniHamed:2008yf}, it is
anticipated that the result of~\cite{Cheung:2015cba} carries over to the
broken phase. However, in the analysis of massive amplitudes we encounter
several technical complications compared to the massless case. In particular,
obtaining shifts with a ``good" large-$z$ behaviour appears to make it
necessary to break manifest little-group covariance by introducing a fixed
spin axis. This requires a careful analysis to ensure that amplitudes for
arbitrary spin configurations can be constructed recursively. Furthermore the
violation of helicity selection rules by mass terms can lead to contributions
to amplitudes with a worse large-$z$ behaviour than in the massless case.

The paper is structured as follows. In Section~\ref{sec:notation} the
little-group covariant spinor formalism for massive
particles~\cite{Arkani-Hamed:2017jhn} is reviewed and related to expressions
for Dirac spinors and polarization vectors for a fixed spin axis. In
Section~\ref{sec:shift} a systematic discussion of shifts of massive momenta
is given, generalizing the extended Risager~\cite{Risager:2005vk} and
BCFW-type shifts used in~\cite{Cheung:2015cba} to the massive case. After
initially constructing shifts of momenta and wave-functions in a little-group
covariant form, a suitable choice of spin axes leads to a similar
$z$-dependence of wave functions as in the massless case. All the required
shifts with two to five legs are constructed explicitly. The large-$z$
behaviour of amplitudes in SBGTs is established in Section~\ref{sec:scaling}.
These results are used in Section~\ref{sec:construct} to identify the minimal
number of shifted lines needed to construct a given amplitude. As an
application, in Section~\ref{sec:select} selection rules for amplitudes with
massive vector bosons are derived using recursion relations and little-group
transformations, providing a new perspective on results obtained using
diagrammatic analysis~\cite{Coradeschi:2012iu} or supersymmetry~\cite{Boels:2011zz}. Details on the
employed spinor conventions and explicit expressions for little-group
transformations changing the spin axis are given in
Appendices~\ref{app:spinors} and~\ref{app:little}, respectively. Recursion
relations with non-lightlike shifts of internal lines are briefly discussed in
Appendix~\ref{app:q2neq0}.

\section{Spinor formalism for massive particles}
\label{sec:notation}

In this section the conventions for the momenta and wave-functions of massive
particles are set up, relating the little-group covariant notation introduced
recently in~\cite{Arkani-Hamed:2017jhn} to the conventions used previously for
on-shell recursion relations for massive quarks~\cite{Schwinn:2007ee}.
The use of little-group transformations to relate wave-functions with
different spin quantum numbers is also discussed.

\subsection{Spinor variables for massive momenta}

At the heart of the spinor-helicity method is the observation that the complex
two-by-two matrix $k_{\alpha\dot\alpha}=k_\mu\sigma^\mu_{\alpha\dot\alpha}$
associated to a light-like momentum factorizes into a product of two-component
Weyl spinors in the two inequivalent fundamental representations of $SL(2,\mathbb C)$,
\begin{equation}
 k_{\alpha\dot\alpha}=k_\alpha k_{\dot\alpha},
\end{equation}
where we refer to the spinors $k_\alpha\in D^{(\frac{1}{2},0)}$ as holomorphic
and the spinors in the conjugate representation
$k_{\dot\alpha}\in D^{(0,\frac{1}{2})}$ as anti-holomorphic. Our Weyl-spinor
conventions follow~\cite{Schwinn:2007ee} and are summarized in
Appendix~\ref{app:spinors}. Similarly, a massive momentum satisfying the
on-shell condition
\begin{equation}
 \label{eq:on-shell}
 k_{\alpha\dot\alpha}k^{\dot\alpha\beta}=m^2\delta_\alpha^\beta
\end{equation}
can be expanded in a basis of two holomorphic and two anti-holomorphic
spinors. Various approaches have been followed in the literature, e.g.\
introducing fixed spin vectors~\cite{Kleiss:1985yh} or using helicity eigenstates~\cite{Dittmaier:1998nn}.

A notation that makes the transformation properties of spinor variables under the little group $SU(2)$ of massive
momenta manifest was introduced in~\cite{Arkani-Hamed:2017jhn}. In this
notation, a massive momentum is parameterized as
\begin{align}
 \label{eq:momentum-little}
 k_{\alpha\dot\alpha}&=
 k_{\alpha}^Ik_{\dot{\alpha},I}=k_{\alpha}^Ik_{\dot{\alpha}}^J\varepsilon_{JI},&
 k^{\dot \alpha\alpha}=k^{\dot{\alpha}}_I k^{\alpha, I}
 &=k^{\dot{\alpha}}_I k^{\alpha}_{J}\varepsilon^{IJ},
\end{align}
where the two-component little-group indices $I$ are raised and lowered with
the two-dimensional antisymmetric tensor. Using the same conventions as for
un-dotted Weyl-spinor indices in~\eqref{eq:indices}, these relations read
\begin{align}
 k_\alpha^I &= \varepsilon^{IJ} k_{\alpha,J},&
 k_{\alpha,J} &= k^I_\alpha \varepsilon_{IJ},&
\varepsilon^{IJ} &=\varepsilon_{IJ} =
\left(\begin{array}{cc} 0 & 1\\ -1 & 0 \\
\end{array} \right),
\label{eq:convention-little}
\end{align}
with identical definitions for the anti-holomorphic spinors.

The massive spinor variables can be chosen to satisfy the $SU(2)$ and $SL(2,\mathbb{C})$ covariant normalization
conventions
\begin{align}
\label{eq:norm-spin}
 \braket{k^Ik^J}&=m\varepsilon^{IJ},&
 \sbraket{k_Ik_J}&=-m\varepsilon_{IJ},\\
 \label{eq:norm-little}
 k_\alpha^Ik_{\beta,I}&=m\varepsilon_{\alpha\beta},&
 k_{\dot\alpha}^Ik_{\dot\beta,I}&=m\varepsilon_{\dot\alpha\dot\beta}.
\end{align}
Holomorphic and anti-holomorphic spinors are related by the Dirac equations
\begin{align}
 \label{eq:dirac-little}
 k_{\alpha\dot\alpha} k^{\dot\alpha}_I&=- mk_{\alpha,I},&
 k^{\dot\alpha\alpha} k_{\alpha}^I&=-mk^{\dot\alpha,I} .
\end{align}

In the construction of the complex continuation of scattering
amplitudes we will be forced to break manifest little-group
covariance and fix a particular basis for the decomposition of the
momenta, as in earlier work on on-shell recursion for massive
momenta~\cite{Schwinn:2007ee}. In this reference a fixed, light-like
reference momentum $q$ is used to decompose a massive momentum into a sum of two light-like vectors,
\begin{equation}\label{eq:decompmomentum}
 k^\mu =k^{\flat;\mu}+\frac{m^2 }{2(q \cdot k)}q^{\mu}.
\end{equation}
The associated holomorphic and anti-holomorphic Weyl spinors
$k^\flat_{\alpha}$, $q_\alpha$ and $ k^\flat_{\dot{\alpha}}$, $q_{\dot\alpha}$
provide a particular example of a basis
for the expansion of a massive momentum, which corresponds to the choice
\begin{equation}
 \label{eq:ref-spinors}
\begin{aligned}
 k_\alpha^1 &= k_\alpha^\flat , & k_\alpha^2 &=
 \frac{m}{\braket{k^\flat q}} q_{\alpha}, \\
 k_{\dot \alpha,1} &= k^\flat_{\dot \alpha} ,&
 k_{\dot \alpha,2} &= \frac{m}{\sbraket{qk^\flat}} q_{\dot{\alpha}}
\end{aligned} 
\end{equation}
in the little-group covariant expressions.
External wave-functions of massive particles can be defined as eigenstates of
the corresponding spin operators with respect to the spin vector
\begin{equation}\label{eq:spin-axis}
n_q^{\mu} = \frac{k^\mu}{m} - \frac{m \,q^{\mu}}{q \cdot k}\,,
\end{equation}
as discussed e.g.\ in~\cite{Boels:2011zz}.
\subsection{Little-group transformations}
By definition, little-group
transformations $R\in SU(2)$ of the Weyl spinors,
\begin{align}
 k^I \to k^{'I} &=R^I{}_J k^J, 
 & k_{I}\to k'_{I} &=-R_I{}^J k_{J} ,
\end{align}
leave the momenta~\eqref{eq:momentum-little} invariant. These definitions
hold both for dotted and un-dotted $SL(2,\mathbb{C})$ indices, which have been
suppressed.
 Note that spinors with lower little-group indices transform in
the dual representation with the transformations
\begin{equation}
 ({R^{T}}^{-1})_I{}^J =(\varepsilon^{-1})_{IK} R^K{}_L\varepsilon^{LJ}=-R_I{}^J,
\end{equation}
where indices of the little-group rotations are raised and lowered with the convention~\eqref{eq:convention-little}.

Infinitesimal little-group transformations,
\begin{equation}
 R^I{}_J=\delta^I_J+\omega^I{}_J+\dots,
\end{equation}
are parameterized by three parameters in the symmetric matrix
$\omega^{IJ}=\omega^{JI}$. The action of infinitesimal little-group
transformations on functions of the spinor variables of a momentum $k$ induces
a representation of the Lie algebra of the little group,
\begin{align}
 \delta\varphi(k^I_{\alpha}, k_{\dot\alpha}^I)&=
 \varphi(k^{'I}_{\alpha},k_{\dot\alpha}^{'I})- \varphi(k^I_{\alpha}, k_{\dot\alpha}^I)
  =-\omega^{I}{}_J(J_k)^J{}_I\varphi   +\mathcal{O}(\omega^2) ,              
\end{align}
with the differential operators~\cite{Chen:2011ve,Conde:2016vxs}
\begin{equation}
( J_k)^{J}{}_I
 =-\left( k^J_{\alpha}\frac{\partial}{\partial
  k^I_\alpha}+k_{\dot\alpha}^J\frac{\partial}{\partial
  k_{\dot\alpha}^I}\right).
\end{equation}
For the determination of spin eigenstates for a fixed spin axis it is useful
to form the linear combinations
\begin{subequations}
\label{eq:generators-spin}
\begin{align}
 J_{k}^- &\equiv ( J_k)^{1}{}_{2}=
    \left(-\frac{\braket{k^\flat q}}{m}k^\flat_\alpha\frac{\partial}{\partial
    q_\alpha} +\frac{m}{\sbraket{qk^\flat}}q_{\dot\alpha}\frac{\partial}{\partial
    k^\flat_{\dot\alpha}} \right),\\
 J_{k}^0 &\equiv \frac{1}{2} ((J_k)^1{}_1-(J_k)^2{}_2) =-\frac{1}{2}\left( k^\flat_\alpha\frac{\partial}{\partial
 k^{\flat}_\alpha}-q_\alpha \frac{\partial}{\partial q_\alpha} -
k^\flat_{\dot\alpha}\frac{\partial}{\partial k^{\flat}_{\dot\alpha}}+q_{\dot\alpha} \frac{\partial}{\partial q_{\dot\alpha} }\right),\\
 J_{k}^+&\equiv( J_k)^{2}{}_{1}=\left(-
    \frac{m}{\braket{k^\flat q}}q_\alpha\frac{\partial}{\partial
    k^{\flat}_\alpha} +\frac{\sbraket{qk^\flat }}{m}k^\flat_{\dot\alpha}\frac{\partial}{\partial q_{\dot\alpha}} \right).
\end{align}
\end{subequations}
These operators 
satisfy the commutation relations
\begin{align}
 \lbrack J_k^0,J_k^\pm\rbrack &=\pm J_k^\pm,&
 \lbrack J_k^+,J_k^-\rbrack &=2 J_k^0,
\end{align}
which show that $J_k^\pm$ serve as raising and lowering
operators for the eigenstates of $J_k^0$.

\subsection{Massive fermions}
Solutions to the massive Dirac equation for particle spinors and their conjugates,
\begin{align}
 \label{eq:dirac}
(\fmslash k-m)u(k)&=0, &\bar u(k)(\fmslash k-m)&=0,
\end{align}
can be constructed in the little-group covariant notation as
\begin{align}
 \label{eq:dirac-spinors}
 u^I(k) &= \begin{pmatrix} k_{\alpha}^I\\ -k^{\dot{\alpha},I} \end{pmatrix},&
\bar u_I(k) &= \begin{pmatrix}  k_I^{\alpha},  k_{\dot{\alpha},I} \end{pmatrix}.
\end{align}
Up to different sign conventions these expressions agree with those
of~\cite{Ochirov:2018uyq} where it is shown that they form helicity
eigenstates. They satisfy the conventional
completeness and normalization conditions due to the properties of the
two-component spinors~\eqref{eq:norm-little}:
\begin{align}
 u^I(k)\bar u_I(k)&=
 \begin{pmatrix}
  k_{\alpha}^I k_I^{\beta} & k_{\alpha}^I k_{\dot{\beta},I}\\
  -k^{\dot{\alpha},I} k_I^{\beta}& -k^{\dot{\alpha},I} k_{\dot{\beta},I}
 \end{pmatrix}
 =
 \begin{pmatrix}
  m\delta_\alpha^\beta & k_{\alpha\dot\beta} \\
  k^{\dot{\alpha}\beta} & m\delta^{\dot\alpha}_{\dot\beta}
 \end{pmatrix}
 =\fmslash k+m, \label{eq:dirac-complete}\\
\bar u_I(k)u^J(k)&=\braket{k_Ik^J}-\sbraket{k_Ik^J}=2m\delta_I^J.
\label{eq:dirac-norm}
\end{align}
 The little-group transformation of the Dirac spinors follows from the index positions,
\begin{align}
 \label{eq:little-dirac-cov}
u^I&\to R^I{}_J u^J, & \bar u_I&\to -\bar u_J R_I{}^J.
\end{align}
Using the translation~\eqref{eq:ref-spinors} it is seen that the Dirac
spinors~\eqref{eq:dirac-spinors} are related to expressions for a fixed spin
axis in the conventions of~\cite{Boels:2011zz} by the correspondence
\begin{equation}
 \label{eq:spinors}
 \begin{aligned}
u(k,\tfrac{1}{2}) &=u^2(k)=
 \begin{pmatrix} \frac{m}{\braket{k^\flat q}} q_\alpha\\
  k^{\flat;\dot{\alpha}} \end{pmatrix},&
   u(k,-\tfrac{1}{2}) &=u^1(k)=
 \begin{pmatrix}  k^\flat_{\alpha}\\ -\frac{m}{\sbraket{qk^\flat}}
  q^{\dot{\alpha}} \end{pmatrix}
 \,,\\ 
\bar u(k,\tfrac{1}{2}) &=\bar u_1(k)=
 \begin{pmatrix} - \frac{m}{\braket{k^\flat q}} q^\alpha,
  k^\flat_{\dot{\alpha}} \end{pmatrix} ,&
 \bar u(k,-\tfrac{1}{2}) &=\bar u_2(k)=
 \begin{pmatrix}  k^{\flat;\alpha}, \frac{m}{\sbraket{qk^\flat}}
  q_{\dot{\alpha}} \end{pmatrix}      \,.
\end{aligned}
\end{equation}
Suitable expressions for the corresponding antiparticle spinors are given by
\begin{align}
 v_I(k) &= \begin{pmatrix}  k_{\alpha,I}\\ k^{\dot{\alpha}}_{I} \end{pmatrix},&
 \bar v^I(k) &= \begin{pmatrix}  k^{\alpha,I},  -k_{\dot{\alpha}}^I \end{pmatrix},
\end{align}
which are identified with the expressions for a fixed spin axis as
\begin{equation}
\begin{aligned}               
 v(k,\tfrac{1}{2}) &=v_1(k)=
 \begin{pmatrix} -\frac{m}{\braket{k^\flat q}} q_\alpha\\
  k^{\flat;\dot{\alpha}} \end{pmatrix},&
 v(k,-\tfrac{1}{2}) &=v_2(k)=
 \begin{pmatrix}  k^\flat_{\alpha}\\ \frac{m}{\sbraket{qk^\flat}}
  q^{\dot{\alpha}} \end{pmatrix},\\
 \bar v(k,\tfrac{1}{2}) &=\bar v^2(k)=
 \begin{pmatrix} \frac{m}{\braket{k^\flat q}} q^\alpha,
  k^\flat_{\dot{\alpha}} \end{pmatrix} ,&
 \bar v(k,-\tfrac{1}{2}) &=\bar v^1(k)=
 \begin{pmatrix}  k^{\flat;\alpha}, - \frac{m}{\sbraket{qk^\flat}} q_{\dot{\alpha}} \end{pmatrix}.
\end{aligned}
\end{equation}
The spinors $\bar u(k,s) $ and $v(k,s)$ describe outgoing particles and
antiparticles with spin quantum number $s$, while $u(k,s)$ and $\bar v(k,s)$
describe incoming particles and antiparticles with reversed spin label.
For the Weyl-spinor conventions of Appendix~\ref{app:spinors} the behaviour of
spinors under a reversal of the momentum is given by
\begin{align}
 \label{eq:crossing}
 v(-k,s)&=\ii\, \text{sgn} (k^0+k^3)\, u(k,s),
 &\bar v(-k,s)=\ii\, \text{sgn} (k^0+k^3)\, \bar u(k,s) . 
\end{align}

A change of the spin axis corresponds to a little-group rotation of the
Dirac spinors,
\begin{align}
 u'(s')&=\sum_{s=\pm\tfrac{1}{2}}\mathcal{R}^{(\frac{1}{2})}_{s's}\, u(s), &
 \bar u'(s')&=\sum_{s=\pm\tfrac{1}{2}} \bar
        u(s)\,\mathcal{R}^{(\frac{1}{2})-1}_{ss'}
\label{eq:little-dirac-spin}        
\end{align}
where the matrix $\mathcal{R}^{(\frac{1}{2})}$ is
given
explicitly in~\eqref{eq:little-dirac} in Appendix~\ref{app:little}.

The Dirac spinors are eigenstates of the generators
$J^0$,
\begin{align}
  J_{k}^0 u(k,s)&=s\, u(k,s), &
  J_{k}^0 v(k,s)&=s v(k,s), 
 \end{align}
and related by the action of $J^\pm$ in~\eqref{eq:generators-spin},
\begin{align}
  J_{k}^\pm u(k,\pm\tfrac{1}{2})&=0,&
                    J_{k}^\pm u(k,\mp\tfrac{1}{2})&=-u(k,\pm\tfrac{1}{2}),\nonumber\\
  J_{k}^\pm v(k,\pm\tfrac{1}{2})&=0,&
  J_{k}^\pm v(k,\mp\tfrac{1}{2})&=v(k,\pm\tfrac{1}{2}),
 \end{align}
which are the expected relations for angular-momentum ladder operators up to a
non-standard phase convention for the
particle spinors.

\subsection{Massive vector bosons}
Polarization vectors of massive spin one particles transform under the
three-dimensional representation of the little group and can be described by symmetric bi-spinors,
\begin{equation}
 \label{eq:pol-tensor}
 \epsilon^{(IJ)}_{\alpha\dot\alpha}(k)=\frac{1}{\sqrt 2m} k^{(I}_\alpha k^{J)}_{\dot\alpha}=\frac{1}{\sqrt 2m} \left(k^{I}_\alpha k^J_{\dot\alpha}+k^{J}_\alpha k^I_{\dot\alpha}\right).
\end{equation}
These satisfy the transversality, orthonormality, and completeness relations
\begin{align}
 k^{\dot\alpha\alpha}
 \epsilon^{(IJ)}_{\alpha\dot\alpha}&
 =0 ,\label{eq:pol-trans} \\
 \epsilon^{(IJ)}_{\alpha\dot\alpha} \epsilon^{(KL)\dot\alpha\alpha}
 & =\left(\varepsilon^{IL}\varepsilon^{JK}+\varepsilon^{JL}\varepsilon^{IK}\right) ,\label{eq:pol-norm}\\
 \epsilon^{(IJ)}_{\alpha\dot\alpha} \epsilon_{(IJ)}^{\dot\beta\beta}
 & =\left(\delta_\alpha^\beta\delta_{\dot\alpha}^{\dot\beta}-\frac{k_{\alpha}^{\dot\beta}k_{\dot\alpha}^\beta}{m^2}\right).
\label{eq:pol-complete}
\end{align}
The polarization vectors transform under little-group transformations as
second-rank tensors, 
\begin{equation}
 \label{eq:little-vector-cov}
 \epsilon^{(IJ)}\to R^I{}_{M} R^J{}_N  \epsilon^{(MN)}.
\end{equation}

The expressions in the spinor formalism for the fixed spin axis (see e.g.~\cite{Dittmaier:1998nn,Boels:2011zz}),
\begin{align}
 \label{eq:pol}
\epsilon_{\alpha\dot{\alpha}}(k,+)& =\sqrt{2} \frac{q_{\alpha}k^\flat_{\dot{\alpha}}}{\braket{qk^\flat}},&
 \epsilon_{\alpha\dot{\alpha}}(k,-) &= \sqrt{2} \frac{k^\flat_\alpha q_{\dot{\alpha}}}{\sbraket{k^\flat q}},&
 \epsilon_{\alpha\dot{\alpha}}(k,0) &= \frac{1}{m}\left(k^\flat_{\alpha} k^\flat_{\dot{\alpha}} - \frac{m^2}{2 q\cdot k } q_{\alpha} q_{\dot{\alpha}}\right),
\end{align}
are related to the little-group covariant notation by the
identifications 
\begin{equation}
 \epsilon(k,+)=\epsilon^{(22)}(k),\quad
 \epsilon(k,-)=-\epsilon^{(11)}(k),\quad
 \epsilon(k,0)=-\sqrt 2\epsilon^{(12)}(k),
\end{equation}
where the minus sign in the definition of $\epsilon(k,-)$ 
ensures the conventional normalization condition
\begin{equation}
 \frac{1}{2}\epsilon_{\alpha\dot\alpha}(\lambda)\epsilon^{\dot\alpha\alpha}(-\lambda')=\epsilon(\lambda)\cdot\epsilon(-\lambda')=-\delta_{\lambda,\lambda'}. 
\end{equation} 
As a result of these conventions, the action of the generators~\eqref{eq:generators-spin} on the polarization
vectors is given by
\begin{equation}
\label{eq:j-eps}
\begin{aligned}
  J_{k}^0 \epsilon(k,s)&=s \epsilon(k,s), \\
J_{k}^\pm \epsilon(k,\pm)&=0 ,\qquad J_{k}^\mp \epsilon(k,\pm)=\pm \sqrt 2 \epsilon(k,0) ,\qquad J_{k}^\pm \epsilon(k,0)=\pm \sqrt 2 \epsilon(k,\pm).
\end{aligned}
\end{equation}
A change of the spin axis is represented as a little-group transformation,
\begin{equation}
 \label{eq:little-vector-spin}
 \epsilon'(s')=\sum_{s=+,0,-}\mathcal{R}^{(1)}_{s's} \epsilon(s),
\end{equation}
where the matrix $\mathcal{R}^{(1)}$ is given in~\eqref{eq:little-vector}.

\section{Complex continuation of massive amplitudes}
\label{sec:shift}

The basis of the proof~\cite{Britto:2005fq} of the original on-shell recursion relation~\cite{Britto:2004ap} and its
generalizations~\cite{Risager:2005vk,Cohen:2010mi,Cheung:2015cba} is the
construction of a complex continuation $A(z)$ of scattering amplitudes
obtained by a deformation of a subset $\mathcal{S}=\{k_i\}$, $i=1,\dots h$ of the four-momenta
of the external particles, parameterized by a complex parameter $z$,
\begin{equation}
k_i\to \hat
k_i(z) \quad\text{for}\quad k_i\in \mathcal{S} ,
\end{equation}
so that the physical
 amplitude is given by $A(0)$.
We consider deformations with the following properties:
\begin{itemize}
\item Four-momenta are deformed by a linear shift in $z$,
 \begin{equation}
  \label{eq:shift-lin}
  \hat k_i(z)=k_i+ z \delta k_i.
 \end{equation}
\item The shift does not modify the mass-shell condition of the external
 momenta, 
 \begin{equation}
  \label{eq:shift-os}
  \hat k_i^2(z)=k_i^2=m_i^2.
 \end{equation}
\item The deformed momenta satisfy momentum conservation if the original momenta do
 so, i.e.
 \begin{equation}
  \label{eq:shift-cons}
  \sum_{\mathcal{S}}\hat k_i(z)=\sum_{\mathcal{S}}k_i,
 \end{equation}
 where all momenta are taken as outgoing.
This condition must only hold for the sum over the full set of shifted
momenta $\mathcal{S}$, i.e.\ there should be no subset
where~\eqref{eq:shift-cons} is satisfied on its own.
 \item For all possible ``factorization channels", i.e.\ all decompositions of
  the set of external momenta $\mathcal{P}$ into
  two subsets $\mathcal{P}=\mathcal{F}\cup\mathcal{F}'$, the sum over
  momenta in each subset is deformed by a light-like vector,
  \begin{equation}
   \label{eq:shift-light}
   \hat K_{\mathcal{F}}(z)=\sum_{\mathcal{F}} \hat k_i(z)=K_{\mathcal{F}} +
   z Q_{\mathcal{F}},\quad Q_{\mathcal{F}}^2=0,
  \end{equation}
  with
  \begin{equation}
   Q_{\mathcal{F}}=\sum_{\mathcal{S}\cap\mathcal{F}} \delta k_i=-\sum_{\mathcal{S}\cap\mathcal{F}'}\delta k_i.
  \end{equation}
  This condition is not necessary (see e.g.~\cite{Cheung:2015cba}) but chosen
  here to simplify the discussion. Shifts with $Q_{\mathcal{F}}^2\neq 0$ do
  not lead to advantages for our purposes, as discussed in Appendix~\ref{app:q2neq0}.
 \end{itemize}
Since poles of Feynman diagrams arise solely through propagators, only simple
poles in $z$ appear for Born amplitudes for the above properties of the shift. 
 The integral of the function
$A(z)/z$ over a circle with $|z|\to \infty$ is given by the sum of the
residues at the poles, $z_{\mathcal{F}}$ and the residue at $z=0$, which gives
the physical amplitude,
\begin{equation}
\label{eq:cauchy} 
\frac{1}{2\pi\ii} \oint \frac{A(z)}{z} = A(0)+\sum_{\text{poles}\, z_\mathcal{F}} 
\text{Res}_{z_\mathcal{F}}\frac{A(z)}{z}.
\end{equation}
Factorization properties of Born
amplitudes imply that the $n$-point
amplitude at the poles $z_{\mathcal{F}}$ factorizes into lower-multiplicity
amplitudes according to\footnote{If the internal particle $\Phi_{\mathcal{F}}$ is a
fermion, a convention-dependent phase factor arises since one of the momenta
$K_{\mathcal{F}} $ and $K_{\mathcal{F'}}=-K_{\mathcal{F}} $ corresponds to an
incoming particle line~\cite{Schwinn:2007ee}. Writing the
numerator of a fermion propagator in terms of the completeness relation and
using the convention~\eqref{eq:crossing} one finds for $K_{\mathcal{F}}^0+K_{\mathcal{F}}^3>0$
\begin{equation*}
\fmslash K_{\mathcal{F}}+M_{\mathcal{F}} = \sum_{s}u(K_{\mathcal{F}},-s)\bar u(K_{\mathcal{F}},s)= \sum_{s}(-\ii v(K_{\mathcal{F}'},-s)(\bar u(K_{\mathcal{F}},s)).
\end{equation*}
}
\begin{equation}
 \label{eq:factor-amp}
\lim_{z\to z_{\mathcal{F}} }A(z)=\sum_s
 A_{\mathcal{F}}(\dots \hat{\Phi}^s_{\mathcal{F}}) 
\frac{\ii}{K_{\mathcal{F}}^2+2z K_{\mathcal{F}}\cdot Q_{\mathcal{F}} -M_{\mathcal{F}}^2} 
A_{\mathcal{F'}}(\hat {\Phi}_{\mathcal{F'}}^{-s},\dots),
\end{equation}
where $\Phi_{\mathcal{F}}^s$ denotes a generic particle with spin projection
$s$, momentum $\hat K_{\mathcal{F}}$ and mass $ M_{\mathcal{F}}$. In the
following, massive vector bosons will be denoted by $W^s$, massive fermions
with $s=\pm\frac{1}{2}$ by $\psi^\pm$, and $\phi$ denotes both physical
scalars or would-be Goldstone bosons. The poles of the complex variable $z$
are located at
\begin{equation}
 \label{eq:pole}
 z_{\mathcal{F}}=-\frac{K^2_{\mathcal{F}}-M_{\mathcal{F}}^2}{2K_{\mathcal{F}}\cdot Q_{\mathcal{F}}}.
\end{equation}
Provided the condition
\begin{equation}
\label{eq:z-infty}
 \lim_{z\to\infty}A(z)=0
\end{equation}
holds, the left hand side of~\eqref{eq:cauchy} vanishes and the physical scattering amplitude is expressed in terms of
lower-point on-shell amplitudes,
\begin{equation}
 \label{eq:recurse}
 A(0)=
 \sum_{\mathcal{F}}\sum_{s}A_{\mathcal{F}}(\dots \hat{\Phi}^s_{\mathcal{F}}) 
\frac{\ii}{K_{\mathcal{F}}^2 -M_{\mathcal{F}}^2} 
A_{\mathcal{F'}}(\hat {\Phi}_{\mathcal{F}'}^{-s},\dots).
\end{equation}
On the right-hand side the shifted momenta in the set $\mathcal{S}$ and the
momentum $\hat K_{\mathcal{F}}$ of the internal line are evaluated at the
poles~\eqref{eq:pole}. A prescription for the reference spinors for internal
massive particles is defined in Section~\ref{sec:multi-shift}.

For internal massive vector bosons, gauge invariance implies that the unphysical degrees of freedom,
i.e. the would-be Goldstone bosons and the fourth polarization vector
$\epsilon^\mu(S)=p^\mu/M$ cancel at the pole~\eqref{eq:pole} so that only the
sum over the three physical polarizations needs to be taken in~\eqref{eq:recurse}.

In this section, we derive little-group covariant expressions for the shift of
spinor variables and the wave functions of massive spin one-half fermions and
vector bosons. However, in order to satisfy the condition~\eqref{eq:z-infty}
we will choose a particular spin axis aligned with the
shift~\cite{Schwinn:2007ee,Cohen:2010mi}. Therefore it must be possible to
recover amplitudes for arbitrary spin axes and spin quantum numbers. There are
two ways to achieve this:
\begin{itemize}
\item Construct the amplitude for a fixed spin state for arbitrary and
 independent spin axes $n_{q_i}$ for all particles. The spin-dependence of
 the scattering amplitude enters only through the polarization wave
 functions, while the remaining truncated amplitude is little-group invariant
 since it can be expressed in terms of momenta, Lorentz tensors and Dirac
 matrices. Therefore amplitudes for arbitrary spin
 states can be obtained using the generators
 $J^\pm_{k_i}$~\eqref{eq:generators-spin}, which only act on the
 wave-function of a single leg $i$.
\item Construct the amplitudes for all spin states for a particular fixed
 choice of the spin axes. Results for arbitrary spin axes can be obtained
 as linear combinations of these results using the finite little-group
 transformations of the polarization wave
 functions~\eqref{eq:little-dirac-spin} and~\eqref{eq:little-vector-spin}.
\end{itemize}

In the remainder of this Section we construct all possible $h$-line shifts of
massive particles satisfying the
properties~\eqref{eq:shift-lin}--\eqref{eq:shift-light}.
The large-$z$ behaviour~\eqref{eq:z-infty} is investigated in Section~\ref{sec:scaling} while the ability to construct the
amplitudes for all spin states is discussed in Section~\ref{sec:construct}.

\subsection{Shifts of massive momenta}
\label{sec:shift-momenta}

We first consider the shift of a single massive momentum $k$ such that the
on-shell condition~\eqref{eq:shift-os} remains satisfied.
In the little-group covariant notation, the shift can be defined by introducing
pairs of holomorphic and anti-holomorphic spinors $\eta_\alpha^I$ and
$\eta_{\dot \alpha,I}$ and deforming the spinor variables according to
\begin{align}
 k_\alpha^I&\to \hat k_\alpha^I(z) =k_\alpha^I+z\eta_\alpha^I, &
 k_{\dot\alpha,I}&\to \hat k_{\dot\alpha,I}(z)=k_{\dot\alpha,I}+z \eta_{\dot\alpha,I}.
\end{align}
The on-shell condition can be satisfied by demanding that the normalization
conditions~\eqref{eq:norm-spin} are not modified by the shift,
\begin{align}
 \braket{\hat k^I(z) \hat k^J(z)}&\overset{!}{=}m\varepsilon^{IJ},&
 \sbraket{\hat k_I(z)\hat k_J(z)}&\overset{!}{=}-m\varepsilon_{IJ}.
\end{align}
This implies that the
holomorphic spinors $\eta^I_{\alpha}$ must satisfy the conditions
\begin{align}
 \braket{\eta^I\eta^J}&=\frac{1}{2}\braket{\eta^I \eta_I} \varepsilon^{IJ}=0,&
 \braket{k^{[I}\eta^{J]}}&=\braket{k^I \eta_I} \varepsilon^{IJ}=0,
\end{align}
where it was used that any two-dimensional antisymmetric tensor is
proportional to the totally antisymmetric symbol.
According to the first condition the shift vector factorizes in terms of a
light-like Weyl spinor $\eta_\alpha$ and a little-group spinor $n^I$,
\begin{equation}
 \eta^I_\alpha = n^I \eta_\alpha.
\end{equation}
The second condition then becomes
\begin{equation}
\varepsilon_{IJ}n^I\braket{k^J\eta}=0 ,
\end{equation}
which determines the little-group spinor up to a constant $c$,
\begin{equation}
 n^I=c \braket{\eta k^I}.
\end{equation}
The shift of the anti-holomorphic variable is treated analogously.
The general shift of the massive spinor variables is therefore of the form
\begin{align}
 \label{eq:shift-spinors}
 \hat k_\alpha^I(z) &=k_\alpha^I+zc\eta_\alpha\braket{\eta k^I}, &
 \hat k_{\dot\alpha,I}(z)&=k_{\dot\alpha,I}+zd\eta_{\dot \alpha}\sbraket{k_I\eta}.
\end{align}
It can be checked that the normalization~\eqref{eq:norm-little} is
automatically satisfied by the shifted spinors as well,
\begin{align}
 \hat k_\alpha^I(z)\hat k_{\beta,I}(z)&=m\varepsilon_{\alpha\beta},&
 \hat k_{\dot\alpha}^I(z)\hat k_{\dot\beta,I}(z)&=m\varepsilon_{\dot\alpha\dot\beta}.
\end{align}
This follows from identities such as
$\braket{\eta k^I}k_{\beta,I}=m\eta_\beta$ that result
from~\eqref{eq:norm-little}. Note that within the spin-axis formalism the
shift~\eqref{eq:shift-spinors} corresponds to shifting both the momentum
spinors $k^\flat$ and the reference spinors $q$.

The requirement~\eqref{eq:shift-lin} of a linear shift of the momentum allows
only a shift of the holomorphic or anti-holomorphic spinors alone, so there are
two possible shifts,\footnote{The only other way to avoid a quadratic term in
 $z$ is to fix the shift
 spinors in terms of the momentum spinors such that $\braket{\eta
  k^I}\sbraket{k_I\eta}=0$, e.g. $\ket{\eta}\propto\ket{k^2}$ and
 $\sket{\eta}\propto\sket{k_1}$. The resulting momentum shift
 $\hat k_{\alpha\dot\alpha}(z)=k_{\alpha\dot\alpha}+z\tilde c\,k^2_\alpha
 k_{\dot\alpha,1}$ can be viewed as a special case of both the
 generic holomorphic and anti-holomorphic shifts and need not be considered separately.}
\begin{align}
 \hat
 k^{\rh}_{\alpha\dot\alpha}(z)&=k_{\alpha\dot\alpha}+zc\,\eta_\alpha\braket{\eta k^I}
 k_{\dot\alpha,I}=k_{\alpha\dot\alpha}+zc\,\eta_\alpha(\bra{\eta} \fmslash k)_{\dot\alpha},\label{eq:h-shift-momentum}\\
 \hat
 k^{\ra}_{\alpha\dot\alpha}(z)&=k_{\alpha\dot\alpha}+zd\, k^I_\alpha\sbraket{k_I\eta}
 \eta_{\dot\alpha}=k_{\alpha\dot\alpha}+zd\, (\fmslash k\sket{\eta})_{\alpha}
 \eta_{\dot\alpha}.\label{eq:a-shift-momentum}
\end{align}
The second form of the shifts follows from the decomposition of the
momentum~\eqref{eq:momentum-little} and makes it clear that these results
reproduce the known expressions for the shift of a massive momentum by a
light-like vector~\cite{Badger:2005zh,Cohen:2010mi}. In this form it is also
straightforward to take the massless limit by rescaling
$c\to c/\!\braket{\eta k}$ and $d\to d/\!\sbraket{k\eta}$ so one obtains the
familiar result
\begin{align}
 \hat k^{\rh}_\alpha(z) &=k_\alpha+zc\,\eta_\alpha, &
\hat k^{\ra}_{\dot\alpha}(z)&=k_{\dot\alpha}+zd\,\eta_{\dot \alpha}.
\end{align}
 For shifts of multiple
external momenta discussed in Section~\ref{sec:multi-shift}, the constants $c$
and $d$ need to be chosen such that the condition of momentum
conservation~\eqref{eq:shift-cons} is satisfied.
\subsection{Shifts of wave-functions}
\label{sec:shift-wave}

Expressions for the Dirac spinors~\eqref{eq:dirac-spinors} and polarization
vectors~\eqref{eq:pol-tensor} for a shifted momentum can be easily defined by
replacing the spinors $k^I_\alpha$ and $k_{\dot\alpha,I}$ with the
corresponding shifted quantities~\eqref{eq:shift-spinors}. For the example of
the holomorphic shift one obtains the Dirac wave function
\begin{equation}
 \label{eq:dirac-shift-h}
 \hat u^{I,\rh}(k,z) =
 \begin{pmatrix}  \hat k_{\alpha}^I(z)\\ -k^{\dot{\alpha},I} \end{pmatrix}
=u^I(k)+ zc\braket{\eta k^I}\begin{pmatrix}  \eta_{\alpha}\\
 0 \end{pmatrix}.
\end{equation}
The shifts of the conjugate and anti-particle spinors are defined in complete
analogy. By construction, these spinors satisfy the appropriate equations of
motion~\eqref{eq:dirac}, completeness conditions~\eqref{eq:dirac-complete} and
normalization conditions~\eqref{eq:dirac-norm} for the shifted momentum since
the normalization conventions~\eqref{eq:norm-spin} and~\eqref{eq:norm-little}
are not affected by the shift. Similarly, the polarization vectors
\begin{equation}
  \label{eq:pol-shift-h}
 \hat \epsilon^{(IJ),\rh}_{\alpha\dot\alpha}(k,z)=
 \frac{1}{\sqrt 2m} \hat k^{(I}_\alpha(z) k^{J)}_{\dot\alpha}
 = \epsilon^{(IJ)}_{\alpha\dot\alpha}(k)+ \frac{z c}{\sqrt 2m} \eta_\alpha\braket{\eta k^{(I}} k^{J)}_{\dot\alpha}
\end{equation}
satisfy the transversality condition~\eqref{eq:pol-trans}, the
normalization~\eqref{eq:pol-norm} and the completeness
relation~\eqref{eq:pol-complete} for the shifted momentum.
The corresponding results for the anti-holomorphic shift are
\begin{equation}
 \hat u^{I,\ra}(k,z) = u^{I}(k)-zd\,\sbraket{k_I\eta}
     \begin{pmatrix} 0\\ \eta^{\dot{\alpha}} \end{pmatrix},
\end{equation}
and
\begin{equation}
 \hat \epsilon^{(IJ),\ra}_{\alpha\dot\alpha}(k,z)=
 \frac{1}{\sqrt 2m} k^{(I}_\alpha \hat k^{J)}_{\dot\alpha}(z)
 = \epsilon^{(IJ)}_{\alpha\dot\alpha}(k)+ \frac{z d}{\sqrt 2m} k^{(I}_\alpha \sbraket{k^{J)}\eta}\eta_{\dot\alpha}.
\end{equation}

These results provide little-group covariant expressions for shifted Dirac
spinors and polarization vectors. However, in general the spinors and
polarization vectors for all spin orientations receive a linear shift, which
is not desirable in the discussion of the scaling of the amplitude $A(z)$ for
$z\to\infty$. Nevertheless, the little-group covariant form of the
shift~\eqref{eq:shift-spinors} was recently used for a BCFW shift of one
massive and one massless leg~\cite{Aoude:2019tzn}. In this paper we focus on
purely massive shifts and leave a systematic analysis
of the large-$z$ behaviour of this type of shift for future work.

\subsubsection{Choice of spin axis}
\label{sec:shift-axis}

Using a suitable choice of spin axis aligned with the shift spinors
$\eta_{\alpha},\eta_{\dot\alpha}$ it is possible to simplify
the shifted wave functions to a form similar to the massless
case~\cite{Cohen:2010mi} so that they stay $z$-independent for some spin
quantum numbers.\footnote{ Alternatively, one could consider
 eliminating holomorphic or anti-holomorphic spinor variables using the
 on-shell condition~\eqref{eq:on-shell} and the Dirac
 equation~\eqref{eq:dirac-little} as advocated in~\cite{Arkani-Hamed:2017jhn}
 (see also a related discussion at the Lagrangian
 level~\cite{Chalmers:2001cy}). This corresponds to the introduction of
 higher-dimensional operators and moves the effect of the shift from the wave
 functions to the truncated amplitude. This does not appear to simplify the
 study of the $z\to\infty$ behaviour but may deserve further study. }
To this end, it is useful to choose the 
holomorphic reference spinor in the light-cone decomposition~\eqref{eq:decompmomentum} of a
holomorphically shifted momentum $\hat k_i^{\rh}(z)$ as
\begin{subequations}
 \label{eq:spin-shift}
\begin{equation}
 q_{i,\alpha}= \eta_{i,\alpha},
\end{equation}
 while for an anti-holomorphically shifted momenta $\hat k_j^{\ra}(z)$ the choice
\begin{equation}
 q_{j,\dot\alpha}= \eta_{j,\dot\alpha}
\end{equation}
\end{subequations}
is made. 
In this way only the light-cone projected momentum
spinors are shifted,
\begin{align}
 \hat k^{\flat\rh}_{i,\alpha}(z)&=k^\flat_{i,\alpha}+ z c_i\eta_{i,\alpha},&
 \hat k^{\flat\rh}_{i,\dot\alpha}(z)&=k^\flat_{i,\dot\alpha},\label{eq:h-shift}   \\                   \hat k^{\flat\ra}_{i,\alpha}(z)&=k^\flat_{i,\alpha},&                  
 \hat k^{\flat\ra}_{j,\dot\alpha}(z)&=k^\flat_{j,\dot\alpha}+ z d_j\eta_{j,\dot\alpha}, \label{eq:a-shift}
\end{align}
where the rescaling $c_i\to c_i/\!\braket{\eta_i k_i^\flat}$ and $d_j\to
d_j/\!\sbraket{k_j^\flat\eta_j}$ was performed.
The remaining reference spinors $q_{i,\dot\alpha}$ and $q_{j,\alpha}$
 are still arbitrary at this stage.
With this choice, only the Dirac spinor with negative spin is affected by the
holomorphic shift, as in the massless case,
\begin{align}
 \label{eq:dirac-h}
   \hat u^{\rh}(k_i,\tfrac{1}{2})&=
 \begin{pmatrix}
  \frac{m}{\braket{k_i^\flat \eta_i}}\eta_{i,\alpha}\\
  k_{i,\dot\alpha}^\flat
   \end{pmatrix}  , &
\hat u^{\rh}(k_i,-\tfrac{1}{2})&=
 \begin{pmatrix}
  k^\flat_{i,\alpha}+zc_i\,\eta_{i,\alpha}\\ -\frac{m}{\sbraket{q_ik_i^\flat}}q_{i,\dot\alpha}
 \end{pmatrix}.
\end{align}
In the same way, only the polarization vector with negative spin and the
longitudinal polarization are shifted,
\begin{equation}
 \label{eq:pol-h}
\begin{aligned}
\hat\epsilon^{\rh}_{\alpha\dot{\alpha}}(k_i,+)&=\sqrt{2}\frac{\eta_{i,\alpha}k^\flat_{i,\dot{\alpha}}}{\braket{\eta_i
  k_i^\flat}},\qquad
 \hat \epsilon_{\alpha\dot{\alpha}}^{\rh}(k_i,-) = \sqrt{2} \frac{(k^\flat_{i,\alpha}+z c_i\, \eta_{i,\alpha})q_{i,\dot{\alpha}}}{\sbraket{k_i^\flat q_i}},
\\
 \hat \epsilon^{\rh}_{\alpha\dot{\alpha}}(k_i,0)&=\frac{1}{m}\left((k^\flat_{i,\alpha} + z c_i\,\eta_{i,\alpha}) k^\flat_{i,\dot{\alpha}}-\frac{m^2}{\braket{\eta_i k_i^\flat}\sbraket{k_i^\flat q_i}} \eta_{i,\alpha} q_{i,\dot{\alpha}}\right).
\end{aligned}
\end{equation}
For the anti-holomorphic shift, only the Dirac spinors with positive spin are
affected, 
\begin{align}
 \label{eq:dirac-a}
  \hat u^{\ra}(k_j,\tfrac{1}{2})&=
 \begin{pmatrix}
  \frac{m}{\braket{k_j^\flat q_j}} q_{j,\alpha}\\
  k_{j,\dot\alpha}^\flat+z d_j \,\eta_{j,\dot\alpha}
 \end{pmatrix},&
 \hat u^{\ra}(k_j,-\tfrac{1}{2})&=
 \begin{pmatrix}
  k_{j,\alpha}^\flat\\
  -\frac{m}{\sbraket{\eta_j k_j^\flat}}\eta_{j,\dot\alpha}
 \end{pmatrix},         
\end{align}
while for massive vector bosons again also the longitudinal polarization
vector is shifted in addition to the positive spin,
\begin{equation}
 \label{eq:pol-a}
\begin{aligned}
 \hat\epsilon^{\ra}_{\alpha\dot{\alpha}}(k_j,+)&=\sqrt{2}
 \frac{q_{j,\alpha}(k^\flat_{j,\dot{\alpha}}+ z d_j\,
  \eta_{j,\dot\alpha})}{\braket{q_jk_j^\flat}} ,\qquad
 \hat \epsilon_{\alpha\dot{\alpha}}^{\ra}(k_j,-) = \sqrt{2} \frac{k^\flat_{j,\alpha} \eta_{j,\dot{\alpha}}}{\sbraket{k_j^\flat \eta_j}},\\
 \hat \epsilon^{\ra}_{\alpha\dot{\alpha}}(k_j,0) &=\frac{1}{m}\left(k^\flat_{j,\alpha}( k^\flat_{j,\dot{\alpha}}+ z d_j\, \eta_{j,\dot{\alpha}})-\frac{m^2}{\braket{q_j k_j^\flat}\sbraket{k_j^\flat \eta_j}} q_{j\alpha} \eta_{j,\dot{\alpha}}\right).
\end{aligned}
\end{equation}

\subsection{Multi-line shifts}
\label{sec:multi-shift}

To construct complex deformations of scattering amplitudes, the set $\mathcal{S}$ of
shifted momenta is split into two subsets, $\mathcal{S}=\mathcal{H}\cup
\mathcal{A}$, where a holomorphic shift~\eqref{eq:h-shift-momentum} is
performed for
a subset of momenta $k_i\in \mathcal{H}$ while momenta $k_j\in \mathcal{A}$
are deformed by an anti-holomorphic shift~\eqref{eq:a-shift-momentum}.
These candidate shifts must then be constrained so that the
conditions~\eqref{eq:shift-cons} and~\eqref{eq:shift-light} are satisfied.
Momentum conservation~\eqref{eq:shift-cons} implies the condition
\begin{align}
0=  \sum_{\mathcal{H}} \delta k^{\rh}_{i,\alpha\dot \alpha}+
  \sum_{\mathcal{A}} \delta k^{\ra}_{j,\alpha\dot \alpha}
&=  \sum_{\mathcal{H}} c_i \eta_{i,\alpha}(\bra{\eta_i}\fmslash k_i)_{\dot\alpha}+\sum_{\mathcal{A}} d_j\, (\fmslash k_{j}\sket{\eta_j})_{\alpha}\eta_{j,\dot\alpha}    \nonumber\\
 &=\sum_{\mathcal{H}} c_i \eta_{i,\alpha}
 k_{i,\dot\alpha}^\flat+ \sum_{\mathcal{A}} d_j\, k^\flat_{j,\alpha}\eta_{j,\dot\alpha}    ,
\label{eq:shift-cons-fix} 
\end{align}
where the expression in the second line holds for the choice of reference
spinors~\eqref{eq:spin-shift} and re-scaled coefficients. For generic shift
spinors $\eta_i$, this identity provides four constraints for the $h$
coefficients $c_i$ and $d_j$ so that a solution always exists for $h\geq
4$. For the cases $h=2,3$ solutions can be obtained for special choices of
shift spinors corresponding to massive generalizations
of the BCFW and Risager constructions, as discussed below.

The condition~\eqref{eq:shift-light} implies that the sum of shifted momenta
for every factorization channel $\mathcal{F}$ is light-like, i.e.\ the quantities
\begin{align}
 Q_{\mathcal{F},\alpha\dot\alpha}&=\sum_{\mathcal{H}\cap\mathcal{F}}
 c_i \eta_{i,\alpha}(\bra{\eta_i}\fmslash k_i)_{\dot\alpha}+\sum_{\mathcal{A}\cap\mathcal{F}} d_j\, (\fmslash k_{j}\sket{\eta_j})_{\alpha}\eta_{j,\dot\alpha}   \nonumber\\
 &=\sum_{\mathcal{H}\cap\mathcal{F}} c_i \eta_{i,\alpha}
 k_{i,\dot\alpha}^\flat+ \sum_{\mathcal{A}\cap\mathcal{F}} d_j\, k^\flat_{j,\alpha}\eta_{j,\dot\alpha} 
 \label{eq:shift-light-spinor}
\end{align}
must factorize into a product of two-component spinors, $ Q_{\mathcal{F},\alpha\dot\alpha}= Q_{\mathcal{F},\alpha}Q_{\mathcal{F},\dot\alpha} $, for all choices of $\mathcal{F}$.
Since the number of factorization channels in general exceeds the
number $h-4$ of the remaining free coefficients, this requires special
choices of the shift spinors $\eta_i$.
Analogously to the massless case considered
in~\cite{Cheung:2015cba} this leaves two
possibilities:
 Generalizations of the construction of
 Risager~\cite{Risager:2005vk} where only holomorphic or
 anti-holomorphic shifts are performed or
generalizations of the BCFW construction by performing a
 holomorphic shift of $h-1$ legs and an anti-holomorphic shift for one
 leg, or vice versa. The extension of these constructions to the massive case
 are discussed in Sections~\ref{sec:risager} and~\ref{sec:bcfw},
 respectively. The generalization of the original two-line BCFW shift to the
 massive case needs to be treated separately and is discussed in Section~\ref{sec:2line}.
Shifts where $l>1$ lines are shifted holomorphically and $h-l>1$ lines
anti-holomorphically lead to factorization channels with
$Q_{\mathcal{F}}^2\neq 0$ and are discussed in Appendix~\ref{app:q2neq0}.

In the application of the recursion relation, also the spin states of the
internal particle $\Phi^s_{\mathcal{F}}$ in the factorized
amplitudes~\eqref{eq:factor-amp} need to be defined. 
It is useful to choose the reference spinors in terms of the factorized
spinors of the internal shift~\eqref{eq:shift-light-spinor},
\begin{align}
 q_{\mathcal{F},\alpha}&= Q_{\mathcal{F},\alpha},&
 q_{\mathcal{F},\dot \alpha}&= Q_{\mathcal{F},\dot \alpha},
\label{eq:ref-internal}
\end{align}
which implies the light-cone decomposition
of the internal momentum 
\begin{equation}
 K_{\mathcal{F},\alpha\dot\alpha}=K_{\mathcal{F},\alpha}^\flat
 K_{\mathcal{F},\dot\alpha}^\flat +
 \frac{K_{\mathcal{F}}^2}{2(K_{\mathcal{F}}\cdot Q_{\mathcal{F}})}Q_{\mathcal{F},\alpha}Q_{\mathcal{F},\dot \alpha}.
\end{equation}
This choice results in a simple expression for the shifted internal momentum,
\begin{equation}
 \label{eq:shift-internal}
 \hat K_{\mathcal{F},\alpha\dot\alpha}(z)=K_{\mathcal{F},\alpha}^\flat
 K_{\mathcal{F},\dot\alpha}^\flat +
 \left(\frac{M_{\mathcal{F}}^2}{2(K_{\mathcal{F}}\cdot Q_{\mathcal{F}})}+
  (z-z_{\mathcal F}) \right)Q_{\mathcal{F},\alpha} Q_{\mathcal{F},\dot \alpha},
\end{equation}
which has been expressed in such a way that the on-shell condition,
\begin{equation}
 \hat K_{\mathcal{F}}^2(z_{\mathcal{F}})=M_{\mathcal{F}}^2,
\end{equation}
at the pole position~\eqref{eq:pole} is manifestly satisfied. 
Note the light-cone projected momentum is not affected by the shift,
i.e. $\hat K_{\mathcal{F}}^\flat=K_{\mathcal{F}}^\flat$.
The wave-functions of the internal
particles can then be defined in terms of the reference
spinors~\eqref{eq:ref-internal} and the momentum spinors
$K^\flat_{\mathcal{F}}$.
\subsubsection{Two-line BCFW shifts}
\label{sec:2line}

For a two-line BCFW-type shift of massive lines, a solution to the on-shell
condition~\eqref{eq:shift-os} and momentum
conservation~\eqref{eq:shift-cons-fix} was constructed
in~\cite{Schwinn:2007ee} using the fact that two massive momenta $k_{i/j}$ can
be expressed in terms of two light-like vectors $l_{i/j}$ as
\begin{align}
\label{eq:decompmom}
k_i& = l_i + \alpha_j l_j, 
& 
k_j &= \alpha_i l_i + l_j,
\end{align}
with the coefficients
\begin{align}
 \alpha_j &= \frac{2k_i\cdot k_j-\text{sgn}(2k_i\cdot k_j)\sqrt{\Delta}}{2k_j^2},&
\alpha_i &= \frac{2k_i\cdot k_j-\text{sgn}(2k_i\cdot k_j)\sqrt{\Delta}}{2k_i^2},
\end{align}
and
\begin{equation}
\Delta = \left( 2k_i\cdot k_j \right)^2 - 4k_i^2 k_j^2.
\end{equation}
A two-line shift with the shift spinors 
\begin{align}
 \label{eq:eta-2line}
 \eta_{\alpha}&= l_{j,\alpha}, &
 \eta_{\dot\alpha}&= l_{i,\dot\alpha},
\end{align}
leads to the momentum shifts
\begin{align}
 \label{eq:bcfw2}
  \hat k_{i,\alpha\dot\alpha}^{\rh}(z)&=k_{i,\alpha\dot\alpha}+z\,l_{j,\alpha}l_{i,\dot\alpha},&
 \hat k_{j,\alpha\dot\alpha}^{\ra}(z)&=k_{j,\alpha\dot\alpha}-z\,l_{j,\alpha}l_{i,\dot\alpha},
\end{align}
which manifestly have the desired properties~\eqref{eq:shift-cons}
and~\eqref{eq:shift-light}. This result was recently also obtained from the
little-group point of view in~\cite{Herderschee:2019dmc}. Making the
choice~\eqref{eq:eta-2line} in the results obtained in
Section~\ref{sec:shift-wave} provides shifted Dirac
spinors~\eqref{eq:dirac-shift-h} and polarization
vectors~\eqref{eq:pol-shift-h} for arbitrary little-group frames. However,
eliminating the $z$-dependence in some of the external wave-functions by
fixing the spin axes according to~\eqref{eq:spin-shift},
\begin{align}
q_{i,\alpha}&=\eta_\alpha=l_{j,\alpha}, & q_{j,\dot\alpha}=\eta_{\dot\alpha}=l_{i,\dot\alpha} , 
\end{align}
reproduces the definition of shifted Dirac spinors in~\cite{Schwinn:2007ee}.
Also the prescription~\eqref{eq:ref-internal} for the reference spinors of the
internal line reproduces the choice of~\cite{Schwinn:2007ee},
\begin{align}
 q_{\mathcal{F},\alpha}&=\eta_\alpha=l_{j,\alpha},&
 q_{\mathcal{F},\dot \alpha}&=\eta_{\dot \alpha}= l_{i,\dot\alpha}.
\label{eq:ref-internal-bcfw2}
\end{align}

\subsubsection{Multi-line Risager-type shifts}
\label{sec:risager}
In the Risager-type solution to the light-cone
condition~\eqref{eq:shift-light-spinor} for all factorization channels, all
shifts are either exclusively holomorphic or anti-holomorphic and shift
spinors $\eta_i$ are chosen identical. In the holomorphic case the shift of
the spinors is therefore given by
\begin{align}
 \hat k_{i,\alpha}^{I,\rh}(z) &=k_{i,\alpha}^I+zc_i\eta_\alpha\braket{\eta k_i^I},&
 \hat k_{i,\dot\alpha,I}^{\rh}(z)&=k_{i,\dot\alpha,I},
\end{align}
so that the shift of the momentum of internal propagators factorizes
according to 
\begin{equation}
 \label{eq:light-cone-fact}
   Q_{\mathcal{F},\alpha\dot\alpha}= \eta_{\alpha}\sum_{\mathcal{S}\cap\mathcal{F}} c_i(\bra{\eta}\fmslash k_i)_{\dot\alpha}\equiv \eta_{\alpha} Q_{\mathcal{F},\dot\alpha}.
\end{equation}
Choosing the spin axis according to~\eqref{eq:spin-shift} implies that the
same reference spinor $q_{i,\alpha}=\eta_\alpha$ is used for all shifted
particles.  This simplifies the shifted momentum spinors to
\begin{align}
 \hat k^{\flat\rh}_{i,\alpha}(z)&=k^\flat_{i,\alpha}+ z c_i\eta_\alpha,&
 \hat k^{\flat\rh}_{i,\dot\alpha}(z)&=k^\flat_{i,\dot\alpha},
\label{eq:risager}
\end{align}
and shifted Dirac spinors and polarization vectors
are given by~\eqref{eq:dirac-h} and~\eqref{eq:pol-h}.
The light-cone projection of the internal momentum is defined using the
reference spinors
\begin{align}
 q_{\mathcal{F},\alpha}&=\eta_\alpha,&
 q_{\mathcal{F},\dot \alpha}&= Q_{\mathcal{F},\dot \alpha}=\sum_{\mathcal{S}\cap\mathcal{F}}c_i
               k^\flat_{i,\dot\alpha}.
\label{eq:ref-internal-risager}
\end{align}

The condition of
momentum conservation~\eqref{eq:shift-cons-fix} becomes
\begin{equation}
0= \sum_{\mathcal{S}} c_ik_{i,\dot\alpha}^\flat.
\end{equation}
For three shifted momenta, $i\in \{i_1,i_2,i_3\}$ the Schouten identity implies
the solution
\begin{align}
 c_{i_1}&=\sbraket{k_{i_2}^\flat k_{i_3}^\flat}, & c_{i_2}&=\sbraket{k_{i_3}^\flat k_{i_1}^\flat},&
 c_{i_3}&=\sbraket{k^\flat_{i_1}k^\flat_{i_2}},    
\end{align}
in
complete analogy to the massless case~\cite{Risager:2005vk}.
In general, a system of $h$ equations for the coefficients can be obtained
by contracting with all of the $k_{i,\dot\alpha}^\flat$ spinors of the shifted
legs. However, the system is under-determined due to the Schouten identity. 
In our construction of scattering amplitudes we will require four-line and
five-line shifts, for which solutions can be written as
\begin{align}
 \label{eq:risager4}
 c_{i_1}&=\sbraket{k_{i_2}^\flat k_{i_3}^\flat}, & c_{i_2}&=\sbraket{k_{i_3}^\flat k_{i_1}^\flat}+\sbraket{k_{i_3}^\flat k_{i_4}^\flat},&
 c_{i_3}&=\sbraket{k^\flat_{i_1}k^\flat_{i_2}}  + \sbraket{k^\flat_{i_4}k^\flat_{i_2}}   ,&
 c_{i_4}&=\sbraket{k_{i_2}^\flat k_{i_3}^\flat},
\end{align}
and
\begin{align}
 \label{eq:risager5}
 c_{i_1}&=\sbraket{k_{i_2}^\flat k_{i_3}^\flat}, & c_{i_2}&=\sbraket{k_{i_3}^\flat k_{i_1}^\flat},&
 c_{i_3}&=\sbraket{k^\flat_{i_1}k^\flat_{i_2}}  + \sbraket{k^\flat_{i_4}k^\flat_{i_5}}   ,&
 c_{i_4}&=\sbraket{k_{i_5}^\flat k_{i_3}^\flat}, & 
 c_{i_5}&=\sbraket{k_{i_3}^\flat k_{i_4}^\flat}.
\end{align}
The anti-holomorphic Risager-type shift is given analogously
by~\eqref{eq:a-shift} with the choice of reference spinor
$q_{j,\dot\alpha}=\eta_{\dot\alpha}$ for all shifted lines, and with
corresponding solutions for the coefficients $d_j$.

\subsubsection{Multi-line BCFW-type shifts}
\label{sec:bcfw}
Generalizations of the BCFW construction are obtained by performing an
anti-holomorphic shift for one leg $k_j\in\mathcal{A}$ and holomorphic shifts
of $h-1$ legs $k_i\in\mathcal{H}$, with all shift spinors chosen identical,
\begin{align}
 \hat k_{i,\alpha}^{I,\rh}(z) &=k_{i,\alpha}^{I}+zc_i\eta_\alpha\braket{\eta k_i^I}, &
 \hat k_{i,\dot\alpha,I}^{\rh}(z)&=k_{i,\dot\alpha,I},\nonumber\\
\hat k_{j,\alpha}^{I,\ra}(z)& =k_{j,\alpha}^I, &
\hat  k_{j,\dot\alpha,I}^{\ra}(z)&=k_{j,\dot\alpha,I}+zd_j\eta_{\dot
                \alpha}\sbraket{k_{j,I}\eta}.
\label{eq:bcfw-gen}
\end{align}
The condition of light-like shifts $Q_{\mathcal{F}}$ of the internal momenta~\eqref{eq:shift-light-spinor} requires the choice
\begin{equation}
\eta_\alpha\propto(\fmslash k_j\sket{\eta})_\alpha,
\end{equation}
so that the shift factorizes for all factorization channels,
\begin{equation}
 Q_{\mathcal{F},\alpha\dot\alpha}
= (\fmslash k_j\sket{\eta})_\alpha Q_{\mathcal{F},\dot\alpha}.
\end{equation}
The choice of spin axis~\eqref{eq:spin-shift} implies that all legs in
$\mathcal{H}$ share the same holomorphic reference spinor. After re-scaling the
coefficients $c_i$ and $d_j$, the shift and reference spinors become
\begin{align}
 \label{eq:reference-bcfw}
 \eta_\alpha&=q_{i,\alpha}=k^\flat_{j,\alpha},&
 \eta_{\dot\alpha}&=q_{j,\dot\alpha} .
\end{align}
The anti-holomorphic reference spinors $q_{i,\dot\alpha}$ of the legs in
$\mathcal{H}$ and all reference spinors of particle $j$ are kept arbitrary.
The generalized BCFW shift therefore takes the simple form
\begin{align}
 \hat k_{i,\alpha}^{\flat \rh}(z) &=k_{i,\alpha}^{\flat}+zc_i k_{j,\alpha}^\flat, &
 \hat k_{i,\dot\alpha}^{\flat \rh}(z)&=k^\flat_{i,\dot\alpha},\nonumber\\
\hat k_{j,\alpha}^{\flat \ra}(z)& =k_{j,\alpha}^\flat, &
\hat k_{j,\dot\alpha}^{\flat\ra}(z)&=k_{j,\dot\alpha}^\flat+zd_j q_{j,\dot \alpha}.
 \label{eq:bcfw-h}
\end{align}
The shifted Dirac spinors and polarization
vectors are given by~\eqref{eq:dirac-h} and~\eqref{eq:pol-h} with
$\eta_\alpha=k_{j,\alpha}^\flat$ for the holomorphic lines and
by~\eqref{eq:dirac-a} and~\eqref{eq:pol-a} with
$\eta_{\dot\alpha}=q_{j,\dot\alpha}$ for the anti-holomorphic line. 
According to~\eqref{eq:ref-internal} the light-cone
projection of the internal momentum is defined using the reference spinors
\begin{align}
 q_{\mathcal{F},\alpha}&=k_{j,\alpha}^\flat, &
 q_{\mathcal{F},\dot \alpha}&=Q_{\mathcal{F},\dot \alpha}=\sum_{\mathcal{H}\cap\mathcal{F}}c_i k^\flat_{i,\dot\alpha}+\sum_{\mathcal{A}\cap\mathcal{F}}d_j q_{j,\dot\alpha}.
\label{eq:ref-internal-bcfw}
\end{align}

The condition of momentum conservation~\eqref{eq:shift-cons-fix} reads
\begin{equation}
0= \sum_{\mathcal{H}} c_ik_{i,\dot\alpha}^\flat+d_jq_{j,\dot\alpha}.
\end{equation}
In the case of three shifted momenta, $i\in \{i_1,i_2\}$, a solution is found
with help of the Schouten identity
\begin{align}
 c_{i_1}&=\sbraket{k_{i_2}^\flat q_j}, & c_{i_2}&=\sbraket{q_j k_{i_1}^\flat},&
 d_j&=\sbraket{k^\flat_{i_1}k^\flat_{i_2}} .   
\end{align}
The solutions for four-line (five-line) shifts can be obtained
from the expressions~\eqref{eq:risager4} and~\eqref{eq:risager5} for the Risager
shift by the replacement
$c_{i_4}\to d_j$ and $k_{i_4}\to q_j$ ($c_{i_5}\to d_j$ and $k_{i_5}\to q_j$).

The BCFW-type shift with one holomorphic line $k_i$ and $h-1$ anti-holomorphic
lines $k_j$ can similarly be brought to the form
\begin{align}
 \hat k_{i,\alpha}^{\flat \rh}(z) &=k_{i,\alpha}^{\flat}+zc_i q_{i,\alpha}, &
 \hat k_{i,\dot\alpha}^{\flat \rh}(z)&=k^\flat_{i,\dot\alpha},\nonumber\\
\hat k_{j,\alpha}^{\flat \ra}(z)& =k_{j,\alpha}^\flat , &
\hat k_{j,\dot\alpha}^{\flat\ra}(z)&=k_{j,\dot\alpha}^\flat+zd_j k^\flat_{i,\dot \alpha},
\end{align}
with the shift and reference spinors
 \begin{align}
  \eta_{\alpha}&=q_{i,\alpha},&
  \eta_{\dot \alpha}&=q_{j,\dot\alpha} =k_{i,\dot \alpha}^\flat.
\end{align}
The light-cone decomposition of internal momenta is performed using the
reference spinors
\begin{align}
 q_{\mathcal{F},\alpha}&=Q_{\mathcal{F}, \alpha}=\sum_{\mathcal{H}\cap\mathcal{F}}c_i
               q_{i,\alpha}+\sum_{\mathcal{A}\cap\mathcal{F}}d_j
               k_{j,\alpha},&
 q_{\mathcal{F},\dot \alpha}&=k_{i,\dot \alpha}^\flat.
\end{align}

\section{Large-$z$ behaviour of amplitudes}
\label{sec:scaling}

\begin{figure}[t]
 \centering
 \includegraphics[width=.6\textwidth]{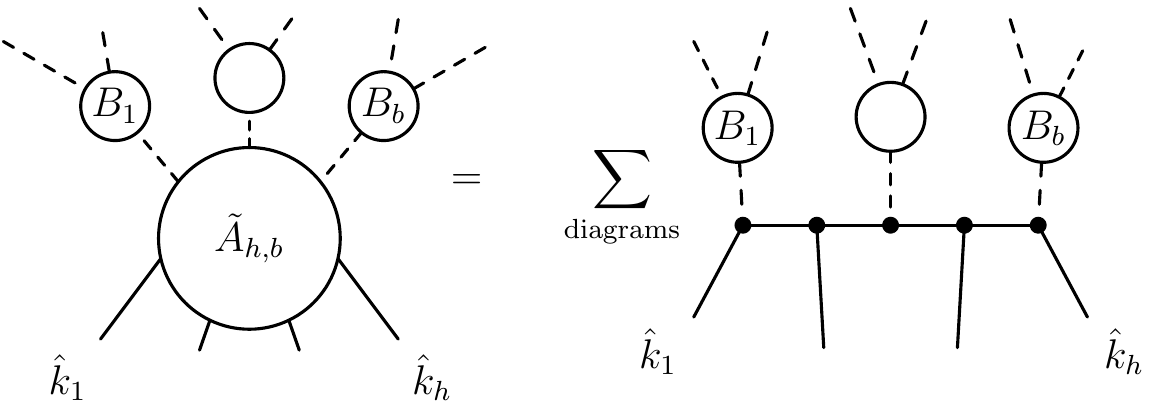}
 \caption{Illustration of the skeleton and background amplitudes 
  appearing in~\eqref{eq:define-skeleton}. The shifted momentum flows
  through solid lines while dashed lines denote background lines.}
 \label{fig:skeleton}
\end{figure}

In this section we obtain bounds on the large-$z$
behaviour of the complex continuation of $n$-point scattering amplitudes, 
\begin{equation}
 \label{eq:def-gamma-amp}
 \lim_{z\to\infty}A_n(z)\sim  z^\gamma,
\end{equation}
under the shifts constructed in Section~\ref{sec:shift}.
Since $z$ enters only through momenta and external wave-functions, which
are all deformed linearly, the exponent $\gamma$ must be an integer, so the
criterion $\gamma<0$ ensures the validity of the condition~\eqref{eq:z-infty}.
The
complex continuation of an $n$-particle scattering amplitude with $h$
shifted external particles can be written in terms of ``skeleton amplitudes"
 $\tilde A_{h,b}(z)$ describing the scattering of the shifted particles,
dressed by insertions of ``background" subdiagrams $B_i$ with the unshifted
external legs,
\begin{equation}
 \label{eq:define-skeleton}
  A_n(z)=  \sum_{b=1}^{n-h}\sum_{\text{diagrams}} \tilde A_{h,b}(z) \prod_{i=1}^{b} B_i ,
 \end{equation}
 see Fig.~\ref{fig:skeleton} for illustration.
More precisely, the background subamplitudes are defined in terms of
off-shell currents $\mathcal{B}_i$, which are given by the off-shell amplitude with $b_i$
external on-shell legs and one off-shell leg with attached propagator.
The numerator of the propagator of the off-shell leg can be written
using a completeness relation in terms of suitably off-shell continued
polarization spinors or vectors and possible additional off-shell terms
(see e.g.~\cite{Schwinn:2005pi}), so that the background currents take the
schematic form
\begin{equation}
 \mathcal{B}_i = \sum_s \chi_i(s) \frac{1}{P_i^2-M^2}
 \chi_i^\dagger(-s)\tilde B_i \equiv \sum_s \chi_i(s) B_i(-s),
\end{equation}
where the sum over $s$ extends beyond the physical polarizations in the
off-shell case.
In~\eqref{eq:define-skeleton} the spin sums are implicit and the polarization
factors $\chi_i$ are included in the definition of the skeleton amplitudes
$\tilde A_{h,b}$, so
that these contain no open spinor or vector indices.
The mass dimension of the background subamplitudes is given
by
\begin{equation}
 [B_i]=
  4-(b_i+1)-2=1-b_i,
\end{equation}
where the term $-2$ arises from the propagator denominator of the
off-shell leg. This split into skeleton and background amplitudes is similar to
the background field analysis in~\cite{Cheung:2015cba}, although the precise
definitions of the background insertions are
somewhat different and lead to different mass dimensions of the skeleton and
background amplitudes compared to our diagrammatic definition.

 The skeleton amplitudes with $h$ shifted lines and $b$ background insertions
 can be broken down into building blocks according to
 \begin{equation}
  \label{eq:amp-skeleton}
 \tilde A_{h,b}(z) =\frac{N_{h,b}(z) }{D_{h,b}(z)}\prod_a g_a
 \prod_{\mathcal{S}_\psi}\hat u(z) \prod_{\mathcal{S}_W}\hat
 \epsilon(z)\prod_{i=1}^b \chi_i,
\end{equation}
where $g_a$ are coupling constants, $D_{h,b}$ is the
product of propagator denominators and $N_{h,b}$ the corresponding numerator
function arising from vertex factors and propagator numerators. The set of shifted fermion and vector boson lines is denoted by $
\mathcal{S}_\psi$ and $\mathcal{S}_W$, respectively. 
 The skeleton amplitude is connected, i.e. all propagators
in $\tilde A_{h,b}$ are $z$-dependent, since the shift is assumed to satisfy
the property that the condition of momentum conservation~\eqref{eq:shift-cons}
does not hold for a subset of the shifted legs.

Dimensional analysis relates the dimension of the scattering amplitude in four
space-time dimensions to the mass dimensions of the objects in the ansatz~\eqref{eq:amp-skeleton}
\begin{equation}
 \label{eq:amp-power}
 [\tilde A_{h,b}]=4-(h+b)= [g]+[N_{h,b}]-[D_{h,b}]+\frac{1}{2}h_{\psi}+\frac{1}{2}b_\psi.
\end{equation}
The term $[g]$ denotes the dimension of the product of coupling constants,
$h_{\Phi}$ denotes the number of shifted external legs of particle type $\Phi$
and $b_{\Phi}$ the corresponding background attachments. Similarly,
the number of holomorphically or anti-holomorphically shifted particles
will be denoted by $h_{\Phi}^{\rh}$ and $h_{\Phi}^{\ra}$.

The scaling exponent of the skeleton amplitudes, $\lim_{z\to\infty}\tilde
A_{h,b}(z)\sim  z^{\gamma_{h,b}}$, can be decomposed as
\begin{equation}
 \label{eq:def-gamma}
 \gamma_{h,b}=\gamma_N-\gamma_D,
\end{equation}
where $\gamma_D$ arises from the propagator denominators and $\gamma_N$ arises
from the flow of $z$ through momentum-dependent numerators of Feynman diagrams
and from external wave functions.
Since the background subamplitudes are $z$-independent, the
behaviour of the full amplitude for $z\to \infty$ is
determined by the worst scaling among the skeleton amplitudes,
\begin{equation}
 \gamma = \max \gamma_{h,b}.
\end{equation}
The criterion $\gamma_{h,b}<0$ is therefore sufficient to establish the condition~\eqref{eq:z-infty}.

\subsection{Scaling of external states and Goldstone-boson equivalence}
\label{sec:gold}

With the choice of spin axis described in
Section~\ref{sec:shift-axis}, the large-z behaviour of the
fermion spinors for holomorphic and anti-holomorphic shifts is given by
\begin{equation}
 \label{eq:scale-fermion}
\begin{aligned}
 \hat u^{\rh}(k,s)&\sim z^{-s+\frac{1}{2}}, \\ \hat u^{\ra}(k,s)&\sim z^{s+\frac{1}{2}},
\end{aligned}
\end{equation}
while the vector-boson polarizations behave as
\begin{equation}
 \label{eq:scale-eps-naive}
\begin{aligned}
 \hat \epsilon^{\rh}(k,-)&\sim z^1, & \hat \epsilon^{\rh}(k,0)&\sim z^1,& \hat \epsilon^{\rh}(k,+)&\sim z^0,\\
\hat \epsilon^{\ra}(k,-)&\sim z^0, &
 \hat \epsilon^{\ra}(k,0)&\sim z^1,
 & \hat \epsilon^{\ra}(k,+)&\sim z^1.
\end{aligned}
\end{equation}
The naive scaling of the polarization vectors for the ``good shifts",
$\hat \epsilon^{\ra}(-)$ and $ \hat \epsilon^{\rh}(+)$, is worse than in the
massless case, where a gauge-dependent $1/z$ pole of polarization vectors
improves the scaling. However, using Ward identities it is possible to
establish the $1/z$ suppression of amplitudes for these polarizations also for
a gauge choice with $z$-independent polarization
vectors~\cite{ArkaniHamed:2008yf}. Similarly, gauge cancellations improve the
behaviour of amplitudes with longitudinal polarization vectors compared to the
naive estimate~\cite{Cohen:2010mi}.

For the case of spontaneously broken gauge invariance, the
relevant Ward identity relates vector-boson to Goldstone-boson amplitudes~\cite{Lee:1977eg,Chanowitz:1985hj},
\begin{equation}
 \label{eq:gb-wi}
 k_\mu A^\mu(W(k),\dots)
= m_W A(\phi(k),\dots).
\end{equation}
Using the observation that the shift of the longitudinal polarization
vector is proportional to the momentum shift,
\begin{equation}
  \hat
  \epsilon^{\rh}_{\alpha\dot{\alpha}}(k,0)-\epsilon_{\alpha\dot{\alpha}}(k,0)=\frac{1}{m}
  c z \eta_\alpha k^\flat_{\dot{\alpha}}
  =\frac{1}{m}\left(\hat k_{\alpha\dot{\alpha}}^{\rh}(z)-k_{\alpha\dot{\alpha}}\right),
 \end{equation}
and similarly for the anti-holomorphic shift, the identity~\eqref{eq:gb-wi} can be used to 
obtain the relation for amplitudes with shifted longitudinal vector bosons,
\begin{equation}
  \label{eq:gb-et}
 \hat \epsilon_\mu (k,0) A^\mu(W(\hat k(z)),\dots)=
A(\phi(\hat k(z)),\dots) + r_{k,\mu} A^\mu(W(\hat k(z)),\dots),
\end{equation}
with 
\begin{equation}
 \label{eq:def-r}
 r_{k,\alpha\dot\alpha}=
 \left(\epsilon_{\alpha\dot\alpha}(k,0)-\frac{k_{\alpha\dot\alpha }}{m}\right)
=- \frac{m^2}{ q\cdot k } q_{\alpha} q_{\dot{\alpha}}.
\end{equation}
The amplitudes on the right-hand side depend on $z$, while the vector $r_k$
does not. In the usual application of the Ward identity~\eqref{eq:gb-wi} in
the context of the Goldstone boson equivalence theorem, the high-energy limit
is taken where $r_k\sim m/E$ so that the term involving the
amputated amplitude $A^\mu(W,\dots)$ is subdominant compared to the Goldstone
boson amplitude. Here we consider the $z\to\infty$ limit where the exact
identity~\eqref{eq:gb-et} cannot be simplified further. This is, however,
sufficient to see that the scaling of the longitudinal polarization
vectors~\eqref{eq:scale-eps-naive} overestimates the $z\to\infty$ behaviour.
Note that the Goldstone-boson and the vector-boson amplitude on the right-hand
side of~\eqref{eq:gb-et} may have a different behaviour for $z\to\infty$ and
the contraction of the remainder $r_k$ with the vector-boson amplitude must be
taken into account in the estimate of the large-$z$ behaviour.

The Ward identity~\eqref{eq:gb-wi} can also be used to 
bound the large-$z$ behaviour of amplitudes with vector
bosons with positive spin projection in the holomorphic shift~\cite{ArkaniHamed:2008yf,Boels:2010mj} by noting that the holomorphic shift of the momentum
can be written in terms of the positive-helicity polarization vector,
\begin{equation}
\hat k_{\alpha\dot{\alpha}}^{\rh}(z)-k_{\alpha\dot{\alpha}}= z \, c
\frac{\sbraket{k^\flat q}}{\sqrt 2} \hat\epsilon^{\rh}_{\alpha\dot{\alpha}}(k,+).
\end{equation}
The application of the Ward identity implies
\begin{equation}
  \label{eq:gb-et-2}
 \hat \epsilon^{\rh}_\mu (k,+) A^\mu(W(\hat k(z)),\dots)
=\frac{1}{z} \frac{\sqrt 2}{c\sbraket{k^\flat q}}\left(m A(\phi( \hat k(z)),\dots) -k_{\mu}A^\mu(W (\hat k(z)),\dots)
\right).
\end{equation}
An analogous identity holds for 
$\hat\epsilon^{\ra}(k,-)$ in the anti-holomorphic shift.

Altogether these results show that gauge
cancellations, as encoded in the Ward identity~\eqref{eq:gb-wi}, improve the
scaling behaviour of massive vector bosons compared to the naive
estimates~\eqref{eq:scale-eps-naive}
so that the effective
scaling is determined by the spin projection $s=0,\pm 1$, 
\begin{equation}
 \label{eq:scale-eps}
\begin{aligned}
 \hat \epsilon^{\rh}(k,s)&\sim z^{-s}, \\
\hat \epsilon^{\ra}(k,s)&\sim z^s.
\end{aligned}
\end{equation}
Therefore the
effective behaviour of amplitudes for the ``good shifts" of vector bosons is
identical to the massless case while the longitudinal vector bosons behave
like scalars, as intuitively anticipated from Goldstone boson equivalence. 
The use of the Ward identity is a key place of our analysis where the
consequences of spontaneously broken gauge invariance are employed. 

\subsection{Propagator scaling}
For internal lines with light-like shifts, every propagator denominator in the
skeleton amplitude is linear in $z$ so that
\begin{equation}
 \label{eq:gamma-d}
 \gamma_D=\frac{1}{2}[D_{h,b}]=d,
\end{equation}
where $d$ is the number of propagators in the skeleton amplitude. This can be
estimated using the topological identities of tree diagrams
\begin{align}
 \sum_nv_n&=d+1,\\
 \sum_n n v_n &=2d +e ,
\end{align}
where $v_n$ is the number of vertices with valency $n$ and $e=h+b$ is the
number of external legs of the skeleton amplitude.
The number of propagators is bounded from above and below in terms of the
smallest and highest valencies $n_{\text{min}}$ and $n_{\text{max}}$, 
\begin{equation}
\frac{h+b-n_{\text{max}}}{n_{\text{max}}-2}\leq\gamma_D\leq \frac{h+b-n_{\text{min}}}{n_{\text{min}}-2}.
\end{equation}
In the following we limit ourselves to a renormalizable SBGT where $n_{\text{min}}=3$ and $n_{\text{max}}=4$ so that
\begin{equation}
 \label{eq:bound-prop}
\frac{h+b-4}{2}\leq\gamma_D\leq h+b-3.
\end{equation}
This condition can easily be relaxed for the interesting application of
on-shell methods in effective-field-theories of SBGTs with higher-dimensional
operators~\cite{Azatov:2016sqh,Shadmi:2018xan,Ma:2019gtx,Aoude:2019tzn,Durieux:2019eor}. In
the derivation of the large-$z$ behaviour in this section, only the upper
bound in~\eqref{eq:bound-prop} enters, so these results are also valid in the
presence of higher-dimensional operators.

\subsection{Bounds for generic shifts with $Q_{\mathcal{F}}^2=0$}
\label{sec:bound-generic}
In addition to the bound on the denominator~\eqref{eq:bound-prop}, the
estimate~\eqref{eq:def-gamma} of the large $z$-scaling of the skeleton
amplitude requires a bound on $\gamma_N$, i.e.\ on the $z$-dependence of the
numerator function $N_{h,b}$ in the ansatz~\eqref{eq:amp-skeleton}, contracted
with the polarization functions of shifted legs and background insertions.
Independent of the structure of the shift, the most conservative estimate is
obtained from the mass dimension of the numerator
function~\cite{Cheung:2015cba}, which can be expressed in terms
of~\eqref{eq:amp-power}. Adding the scaling of the external
fermions~\eqref{eq:scale-fermion} and the improved estimate for vector
bosons~\eqref{eq:scale-eps} gives the bound
\begin{align}
\label{eq:numerator-conservative}
 \gamma_N&\leq
 [N_{h,b}] 
  + \sum_{\mathcal{H}_\psi} \left(-s_i +\frac{1}{2}\right)+\sum_{\mathcal{A}_\psi} \left(s_j+\frac{1}{2}\right)- \sum_{\mathcal{H}_W} s_i+\sum_{\mathcal{A}_W} s_j\nonumber\\
&=4-(h+ b)- [g]-\frac{b_\psi}{2}+[D_{h,b}]- \sum_{\mathcal{H}} s_i+\sum_{\mathcal{A}} s_j.
\end{align}
For the concrete examples of the extended Risager and BCFW shifts, better
estimates can be obtained, as discussed below.
Using the estimate of the scaling of the propagator~\eqref{eq:bound-prop}, a
conservative bound for the scaling of the amplitude is obtained,
\begin{align}
 \gamma&\leq 4-(h+ b)- \text{min}[g]-\frac{b_\psi}{2}+\gamma_D- \sum_{\mathcal{H}} s_i+\sum_{\mathcal{A}} s_j\nonumber\\
& \leq 1 -\text{min}[g]- \sum_{\mathcal{H}} s_i+\sum_{\mathcal{A}} s_j,
 \label{eq:bound-generic}
\end{align}
where $ b_\psi\geq 0$ was used. This result agrees with the case
$n_{\text{min}}=3$ in the result for the massless case in Eq. (26)
in~\cite{Cheung:2015cba}.
However, the first line of~\eqref{eq:bound-generic} shows that this bound can
be improved in amplitudes with fermionic background insertions.
Note that the Feynman diagrams contributing to
the amplitude may have different mass dimension $[g]$ of the product of
coupling constants, so the smallest among these values must be taken for the
bound in~\eqref{eq:bound-generic}.

\subsection{Bounds for multi-line Risager and BCFW shifts}
The generic bound~\eqref{eq:bound-generic} can be improved by an analysis of the structure of the
contraction of the numerator function in the skeleton
amplitude~\eqref{eq:amp-skeleton} with the external and background
wavefunctions. Compared to the corresponding discussion of the massless
case~\cite{Cheung:2015cba}, complications for massive particles arise
due to the presence of reference spinors and the need to apply Ward identities
to bound the scaling of amplitudes with vector bosons. The following
analysis does not cover two-line BCFW
shifts, where the estimate~\eqref{eq:bound-generic} can be used.

Using the definition of the polarization wave functions, the numerator can be
written as a polynomial in holomorphic and anti-holomorphic spinor products of
momentum spinors and reference spinors. The $z$-dependence in the terms
contributing to the skeleton amplitude is of the schematic form
\begin{equation}
 \label{eq:skeleton-scale}
 \tilde A_{h,b}(z) \sim R_{h,b} \frac{\braket{\hat
   k^{\flat\rh}_i(z)\, \cdot\,}^{\alpha}
   \sbraket{\hat
   k^{\flat\ra}_j(z)\, \cdot \,}^{\beta} }{D_{h,b}(z)} .
\end{equation}
The different contributions to the shifted holomorphic and anti-holomorphic
spinor products $\braket{\hat k^{\rh}_i(z)\, \cdot\,}$ and
$\sbraket{\hat k^{\ra}_j(z)\, \cdot \,}$ will be analyzed for the extended
Risager and BCFW-type shifts below. The function $ R_{h,b} $ includes
unshifted spinor products, particle masses and coupling
constants.
The analysis will be performed in the 't Hooft-Feynman gauge where the
numerators of vector-boson propagators are momentum independent.
Note that the exponents $\alpha$ and $\beta$ are positive since spinor products in the denominator can only arise from
the definitions of the polarization wave functions, which are not shifted and
therefore included in the remainder function $ R_{h,b} $.

For all the considered shifts the spinor products involving
shifted spinors in the skeleton amplitude~\eqref{eq:skeleton-scale} are linear in $z$ so that the large-$z$ scaling of the
numerator function is given by $\gamma_N=\alpha+\beta$.
However, as discussed in Section~\ref{sec:gold}, Ward identities improve the large-$z$ behaviour for the ``good
shifts" of vector bosons and for longitudinal bosons in SBGTs compared to
naive scaling estimates. Therefore, the estimate of
$\alpha$ and $\beta$ should be based on the right-hand sides
of~\eqref{eq:gb-et} and~\eqref{eq:gb-et-2}. The explicit factor of $1/z$
in~\eqref{eq:gb-et-2} can be taken into account by defining the effective
scaling exponent,
\begin{equation}
 \label{eq:gamma-n}
 \gamma_N=\alpha+\beta-h_{W(+)}^{\rh}-h_{W(-)}^{\ra},
\end{equation}
with the numbers 
$h^{\rh}_{W(+)}$ ($h^{\ra}_{W(-)}$) of holomorphically
(anti-holomorphically) shifted vector bosons with positive (negative) spin
projection.

\subsubsection{Multi-line Risager-type shifts}
In an extended holomorphic Risager-type shift as constructed in
Section~\ref{sec:risager}, the momenta of all particles in the
set~$\mathcal{S}$ are shifted as~\eqref{eq:risager} and the spin axis of all
shifted particles is taken as $ q_{i,\alpha}=\eta_\alpha$. Therefore the only
spinor products that are affected by the shift are
\begin{align}
 \braket{\hat k_i^{\flat\rh}(z) \hat k_k^{\flat\rh}(z)}&= \braket{k_i^\flat k_k^\flat}+z(c_i\braket{\eta k_k^\flat}+c_k\braket{k_i^\flat\eta}),&
\braket{\hat k_i^{\flat\rh}(z) \chi}&= \braket{k_i^\flat\chi}+z c_i\braket{\eta \chi},
\end{align}
where $\chi$ denotes a generic unshifted spinor, e.g.\ from background
insertions. Due to the choice of spin axes of the shifted particles, products
involving the reference spinors,
$ \braket{\hat k_i^{\flat\rh}(z) q_k}= \braket{k_i^\flat \eta}$ are not shifted and
accordingly contribute to the remainder function $R_{h,b}$
in~\eqref{eq:skeleton-scale}. The same holds for all anti-holomorphic spinor
products, so $\beta=0$. Therefore the relevant contributions to the
holomorphic spinor products are of the form
\begin{equation}
 \label{eq:holomorphic-risager}
 \braket{\hat
   k^{\flat\rh}_i(z)\, \cdot\,}^{\alpha}=\braket{\hat
   k^{\flat\rh}_i(z)\hat k^{\flat\rh}_k(z)}^{\alpha_1}\braket{\hat
   k^{\flat\rh}_i(z) \chi }^{\alpha_2} .
\end{equation}
The most conservative bound is
obtained by assuming that the function $ R_{h,b}$ does not
contain any holomorphic spinors. Since spinor products of
shifted holomorphic spinors are linear in $z$, the upper bound on $\gamma_N$
is obtained from half of the number of those
holomorphic spinors that contribute to $\alpha_1$ and $\alpha_2$. 
This receives the following contributions:
\begin{itemize}
\item The number of holomorphic spinors in the numerator function
 $N_{h,b}$, which can be bound by the mass dimension $[N_{h,b}]$ since
 every four-momentum in the numerator gives rise to one holomorphic spinor.
\item The number of shifted external holomorphic
 spinors $\hat k_{i,\alpha}^{\flat\rh}(z)$, which arise only from the wave
 functions $\hat u_i^{\rh}(-\tfrac{1}{2})$ and $\hat \epsilon_i^{\rh}(-)$ after applying the Ward identity.
\item The number of shifted vector bosons with positive spin projection. This
 contribution arises in addition to the explicit term
 in~\eqref{eq:gamma-n} since the  term $k_{k,\mu}A^\mu(W,\dots)$ in the
 Ward-identity~\eqref{eq:gb-et-2} can give rise to a spinor product
 $\braket{\hat k_i^{\flat\rh}(z)k_k^\flat}$ in the numerator.
 \footnote{No such contribution arises for the longitudinal gauge bosons,
  since the vector $r_k$ in the Ward identity~\eqref{eq:gb-et}
 is proportional to the reference spinors $\eta_\alpha
 q_{k,\dot\alpha}$ and does not contribute to $\alpha$.}
\item The number $b_\psi+b_W$ of effective polarization functions $\chi$ of the
 background legs, which contain at most one holomorphic spinor in the numerator.
 \end{itemize}
 Including all of these contributions gives rise to the estimate
\begin{align}
 \gamma_N
& \leq\frac{1}{2}\left([N_{h,b}] +b_\psi+b_W
 +h_{\psi(-)}+h_{W(-)}-h_{W(+)}\right)\nonumber\\
 &=\frac{1}{2}\left(4-h-\text{min}[g]-b_\phi-\frac{b_\psi}{2}-\sum_{\mathcal{S}}s_i\right)+\gamma_D,
\label{eq:numerator-risager}  
\end{align}
where~\eqref{eq:amp-power} was used with $b=b_\phi+b_\psi+b_W$. We have also simplified
$h_{\psi(-)}-\frac{1}{2}h_\psi=\frac{1}{2}(h_{\psi(-)}-h_{\psi(+)})=-\sum_{\psi}s_i$. 
Therefore the full large-$z$ behaviour of the amplitude can be bound by
\begin{equation}
 \label{eq:risager-scale}
 \gamma_{\text{Risager}}^{\rh}=\gamma_N-\gamma_D\leq 
\frac{1}{2}(4-h-\text{min}[g]-\sum_{\mathcal{S}}s_i),
\end{equation}
since $b_\psi,b_\phi\geq 0$.
Note that for specific examples the estimate can be improved by taking the
concrete structure of the background into account.
For the anti-holomorphic Risager shift, this bound
holds with the replacement $s_i\to -s_j$. These results agree with those for
the massless case~\cite{Cheung:2015cba} and for massive
all-line shifts~\cite{Cohen:2010mi}.

\subsubsection{Multi-line BCFW-type shifts}
In the generalized BCFW shift~\eqref{eq:bcfw-h}, $h-1$ momenta $\hat
k_i^{\rh}$ are shifted holomorphically and a single momentum $\hat
k_j^{\ra}$ anti-holomorphically with the shift spinors
$\eta_\alpha=k_{j,\alpha}^\flat$ and arbitrary $\eta_{\dot\alpha}=q_{j,\dot\alpha}$.
The choice of the reference spinors for the shifted legs is given in~\eqref{eq:reference-bcfw}.
To estimate the number of spinor products with anti-holomorphically
shifted spinors in~\eqref{eq:skeleton-scale}, note that momentum conservation can be used in the numerator
function $N_{h,b}$ to eliminate $\hat k_j^{\ra}$ in favour of the
remaining external momenta of the skeleton amplitude. Therefore only the
external wave-functions need to be considered. After application of the Ward
identity, only $\hat \epsilon^{\ra}_j(+)$ and $\hat u^{\ra}_j(+\tfrac{1}{2})$ contribute,
\begin{equation}
  \label{eq:bcfw-scale-square}
  \beta\leq h_{\psi(+)}^{\ra}+h_{W(+)}^{\ra}.
\end{equation}

The nontrivial holomorphic spinor products of the shifted spinors among themselves and
with other spinors are\footnote{Here we exclude two-line BCFW
 shifts, where only one
momentum is shifted holomorphically.} 
\begin{align}
 \braket{\hat k_i^{\flat\rh}(z) \hat k_k^{\flat\rh}(z)}&= \braket{k_i^\flat k_k^\flat}+z(c_i\braket{k_j^\flat k_k^\flat}+c_k\braket{k_i^\flat k_j^\flat}),\\
 \braket{\hat k_i^{\flat\rh}(z) \chi}&= \braket{k_i^\flat \chi}+zc_i\braket{k_j^\flat \chi},\\
  \braket{\hat k_i^{\flat \rh}(z) q_j^{\ra}}&=
  \braket{k_i^\flat q_j^{\ra}}+z c_i\braket{k_j^\flat q_j^{\ra}},
\end{align}
while the choice of shift and reference spinors implies that
$\braket{\hat k^{\flat \rh}_i(z)q_k^{\rh}}$ and $\braket{\hat
 k^{\flat \rh}_i(z)\hat k^{\flat \ra}_j}$ are not shifted and in particular $\braket{q_i^{\rh}\hat k^{\flat \ra}_j}=0$.
Therefore the relevant contributions to the holomorphic spinor products
in~\eqref{eq:skeleton-scale} are of the form
\begin{equation}
 \label{eq:holomorphic-bcfw}
 \braket{\hat
   k^{\flat\rh}_i(z)\, \cdot\,}^{\alpha}=\braket{\hat
   k^{\flat\rh}_i(z)\hat k^{\flat\rh}_k(z)}^{\alpha_1}\braket{\hat
   k^{\flat\rh}_i(z) \chi }^{\alpha_2} \braket{\hat
   k^{\flat\rh}_i (z)q_j^{\ra}}^{\alpha_3}.
 \end{equation}
Following the reasoning for the Risager-type shift, $\alpha$ can be bound by
 half the number of the holomorphic
spinors contributing to these spinor products. The single
anti-holomorphically shifted particle can contribute spinors
$k_{j,\alpha}^{\ra}$ and $q_{j,\alpha}^{\ra}$, where
the former drops out of~\eqref{eq:holomorphic-bcfw} while the
letter can contribute to $\alpha_3$.
The number of relevant spinors
receives the following contributions:
\begin{itemize}
\item The number of holomorphic spinors in $N_{h,b}$, the holomorphically
 shifted $\hat u^{\rh}_i(-\tfrac{1}{2})$, $\hat\epsilon^{\rh}_i(-)$ and
 $\hat\epsilon^{\rh}_i(+)$, and the background contributions as in the
 Risager-type shift.
\item The anti-holomorphically shifted polarization vector
 $\hat\epsilon^{\ra}_j(+)$, since it includes a spinor $q^{\ra}_{j,\alpha}$ and therefore
 contributes to $\alpha_3$.
\item For $\hat \epsilon^{\ra}_j(-)$, the term $k_j^\mu A_\mu(W,\dots)$
 in the Ward identity~\eqref{eq:gb-et-2} contains a term involving the reference spinor
 $q^{\ra}_{j,\alpha}$ in the massive case. Similarly, the term involving
 $r_{k_j,\alpha\dot\alpha}\propto q^{\ra}_{j,\alpha}
 \eta_{\dot\alpha}$ in the Ward identity~\eqref{eq:gb-et} contributes for $\hat \epsilon^{\ra}_j(0)$.
\item In the massive case, the spinor $\hat u^{\ra}_j(+\tfrac{1}{2})$
 contains the reference spinor $q^{\ra}_{j,\alpha}$ and contributes one-half
 to $\alpha_3$. However, either only the holomorphic or the anti-holomorphic
 part of a Dirac spinor contributes to a given term in the amplitude. Since
 the possible contribution to $\alpha_3$ is smaller than the contribution of
 the anti-holomorphic component of $\hat u^{\ra}_j(+\tfrac{1}{2})$ to
 $\beta$ already included in~\eqref{eq:bcfw-scale-square}, the former can be
 dropped.
\end{itemize}
Taking all of the contributions to $\alpha$ into account and adding the
explicit powers of $z$ due to the application of the Ward identities gives the
bound 
\begin{align}
 & \alpha-h_{W(+)}^{\rh}-h_{W(-)}^{\ra}\nonumber\\
 &\leq\frac{1}{2}\left([N_{h,b}] +b_\psi+b_W
                +h_{\psi(-)}^{\rh}+h_{W(-)}^{\rh}-h_{W(+)}^{\rh}
+h_{W(+)}^{\ra}+h_{W(0)}^{\ra}-h_{W(-)}^{\ra}\right)
  \nonumber \\
 &=\frac{1}{2}\left(4-h-\text{min}[g]-b_\phi-\frac{b_\psi}{2}
  -\sum_{\mathcal{H}}s_i +\sum_{\mathcal{A}_{W}}s_j-\frac{h_\psi^{\ra}}{2}+h_{W(0)}^{\ra} \right) +\gamma_D ,
\end{align}
where the expression for the dimension of the skeleton
amplitude~\eqref{eq:amp-power} was used and the spin sums were introduced as
discussed above after~\eqref{eq:numerator-risager}. Adding the contribution
from the anti-holomorphic spinors~\eqref{eq:bcfw-scale-square}, the bound on
the scaling of the amplitude can be written in a form similar to the bound for
the Risager shift and an additional contribution depending on the spin of the
single anti-holomorphically shifted particle,
\begin{align}
 \label{eq:bound-bcfw-hol}
 \gamma^{\rh}_{\text{BCFW}}&=\gamma_N-\gamma_D
    \leq \frac{1}{2}\left(4-h-\text{min}[g]-b_\phi-\frac{b_\psi}{2} -\sum_{\mathcal{S}}s_i \right)+     
 \begin{cases}
       2 s_j & W^{+,\ra}, \psi^{+,\ra}\\
      \frac{1}{2}, & W^{0,\ra}\\
      s_j , & W^{-,\ra}, \psi^{-,\ra}
     \end{cases}
\end{align}
For a dominantly anti-holomorphic BCFW-type shift one obtains analogously
\begin{align}
 \gamma^{\ra}_{\text{BCFW}}&
    \leq \frac{1}{2}\left(4-h-\text{min}[g]-b_\phi-\frac{b_\psi}{2} +\sum_{\mathcal{S}}s_i \right)+     
 \begin{cases}
       -2 s_j & W^{-,\ra}, \psi^{-,\ra}\\
      \frac{1}{2}, & W^{0,\ra}\\
      -s_j , & W^{+,\ra}, \psi^{+,\ra}
     \end{cases}
\end{align}

The corresponding result for holomorphic shift in the massless case~\cite{Cheung:2015cba} is
\begin{align}
 \label{eq:bound-bcfw-m0}
 \gamma_{\text{BCFW}}^{\rh,m=0}
    \leq \frac{1}{2}\left(4-h-\text{min}[g]- \sum_{\mathcal{S}}s_i \right)+ 2s^{\ra}.
\end{align}
The better behaviour for the ``good shifts" in the massless case can be
understood as follows:
\begin{itemize}
\item For $\psi^{-,\ra}$ one can argue in the massless case that the change from
 the holomorphic to the anti-holomorphic shift improves the behaviour by one
 power of $z$. In the massive case the spinor $\hat u_i^{\rh}(-\tfrac{1}{2})$ also contains 
 a $z$-independent anti-holomorphic component so not all contributions to the
 amplitude are improved.
\item For $W^{-,\ra}$ the Ward identity~\eqref{eq:gb-et-2} gives rise to a
 contribution involving the reference spinor that does not appear in the
 massless case. 
\end{itemize}

\subsection{Examples}
\label{sec:bound-example}

We will illustrate the bounds~\eqref{eq:risager-scale}
and~\eqref{eq:bound-bcfw-hol} for simple examples in order to illuminate
differences of the massive and massless cases more concretely. As a first
example, consider a mostly holomorphic four-line BCFW-type
shift~\eqref{eq:bcfw-h},
\begin{align}
 \label{eq:bcfw-4}
 \hat k_{i,\alpha}^{\flat \rh}(z) &=k_{i,\alpha}^{\flat}+zc_i k_{4,\alpha}^\flat, &
 \hat k_{4,\dot\alpha}^{\flat\ra}(z)&=k_{4,\dot\alpha}^\flat+zd_4 q_{4,\dot \alpha},
\end{align}
for the amplitude
\begin{equation}
A_4(\bar \psi_1^{-,\rh}\!\!,\phi_2^{\rh},\phi_3^{\rh},\psi_4^{-,\ra}) 
=\parbox{.18\textwidth}{
\includegraphics[width=.18\textwidth]{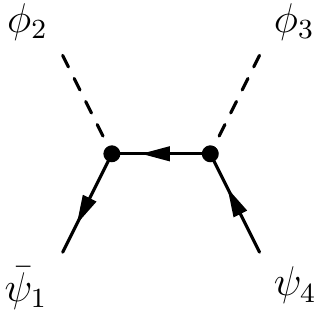}
}
+
\parbox{.18\textwidth}{
\includegraphics[width=.18\textwidth]{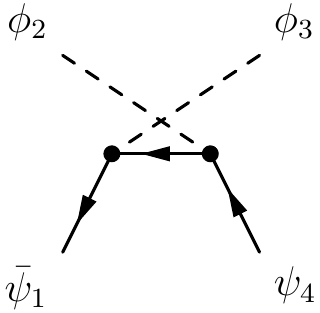}
}
+\parbox{.18\textwidth}{
\includegraphics[width=.18\textwidth]{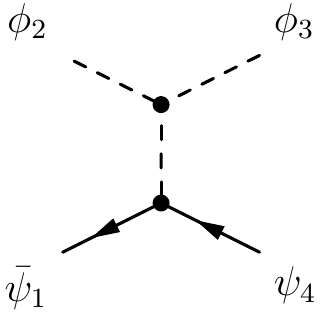}
 }
\end{equation}
in Yukawa theory. In the massless case, according to the
bound~\eqref{eq:bound-bcfw-m0} the shift is allowed with
$ \gamma_{\text{BCFW}}^{\rh,m=0}\leq -\frac{1}{2}$ whereas in the massive case
the estimate~\eqref{eq:bound-bcfw-hol} gives
$\gamma_{\text{BCFW}}^{\rh}\leq 0$. For massless fermions, due to helicity
selection rules, the only contribution to the amplitude arises from the
diagram with a triple-scalar vertex,
\begin{equation}
\parbox{.18\textwidth}{
\includegraphics[width=.18\textwidth]{fig-4}
 }\propto
\hat{\bar u}^{\rh}(k_1,-\tfrac{1}{2})\hat v^{\ra}(k_4,-\tfrac{1}{2})
\frac{\mathrm i}{(\hat k_2^{\rh}(z)+\hat k^{\rh}_3(z))^2-M_\phi^2}\sim \frac{1}{z},
\end{equation}
since the shift drops out of the product of the Dirac
spinors~\eqref{eq:dirac-h} and~\eqref{eq:dirac-a}, 
\begin{equation}
 \hat{\bar u}^{\rh}(k_1,-\tfrac{1}{2})\hat v^{\ra}(k_4,-\tfrac{1}{2})=
 \braket{\hat k_1^{\flat \rh}(z)k_4^\flat}=\braket{k_1^\flat k_4^\flat} \sim z^0.
\end{equation}
In the massive case, there is a non-vanishing contribution from the fermion
exchange diagrams, which include a $z$-dependent contribution in the
numerator, for example
\begin{align}
\parbox{.18\textwidth}{
\includegraphics[width=.18\textwidth]{fig-2}
 }
&\propto
 \hat{\bar u}^{\rh}(k_1,-\tfrac{1}{2}) \frac{\hat {\fmslash k}_3\!\!{}^{\rh}(z)+\hat {\fmslash k}_4\!\!{}^{\ra}(z)+m_\psi}{(\hat k_3^{\rh}(z)+\hat k^{\ra}_4(z))^2-m_\psi^2}\hat v^{\ra}(k_4,-\tfrac{1}{2}) \nonumber\\
& \underset{z\to\infty}{\sim} \frac{\bra{ k_1^{\flat \rh}(z)}\hat {\fmslash k}_3\!\!{}^{\rh}(z)\sket{q_4}m_\psi}{(\hat k_3^{\rh}(z)+\hat  k^{\ra}_4(z))^2-m_\psi^2}  
 \sim \frac{z(c_1\braket{ k_4^\flat k^\flat_3}+c_3\braket{ k_1^\flat
 k^\flat_4})\sbraket{k_3^\flat q_4}m_\psi}{(\hat k_3^{\rh}(z)+\hat  k^{\ra}_4(z))^2-m_\psi^2} \sim z^0,
\end{align}
so that the massive bound~\eqref{eq:bound-bcfw-hol} is saturated.
This illustrates how mass-suppressed contributions can give rise to a
worse large-$z$ behaviour than in the massless case.

 In this example, this
complication is not relevant in practice since the
amplitude is constructible using an anti-holomorphic Risager shift of
all lines,
\begin{equation}
 \hat k^{\flat\ra}_{i,\dot \alpha}(z)=k^\flat_{i,\dot \alpha}+ z d_i\eta_{\dot\alpha}, 
\end{equation}
with the reference spinors $q_{i,\dot \alpha}=\eta_{\dot\alpha}$. In
agreement with the anti-holomorphic version
of~\eqref{eq:risager-scale} one obtains for the diagram with a fermion propagator
\begin{align}
\parbox{.18\textwidth}{
\includegraphics[width=.18\textwidth]{fig-2}
 }
&\propto
 \hat{\bar u}^{\ra}(k_1,-\tfrac{1}{2}) \frac{\hat {\fmslash
 k}_3\!\!{}^{\ra}(z)+\hat {\fmslash k}_4\!\!{}^{\ra}(z)+m_\psi}{(\hat
 k_3^{\ra}(z)+\hat k^{\ra}_4(z))^2-m_\psi^2}\hat
 v^{\ra}(k_4,-\tfrac{1}{2}) \sim \frac{1}{z},
\end{align}
since in this case the choice of reference spinors ensures that the
shift drops out in the numerator, $\hat {\fmslash
 k}_3\!\!{}^{\ra}(z)\sket{q_4}=\fmslash k_3\sket{\eta}$, and similarly for
spinor chains involving $\sbra{q_1}$. This example also illustrates
that the large-$z$ behaviour of amplitudes with fixed spin quantum numbers
generally depends on the choice of the spin axes, in contrast to the
massless case where the reference spinors are unphysical auxiliary quantities.

As an example for the large-$z$ behaviour of amplitudes with massive vector
bosons for the BCFW-type shift~\eqref{eq:bcfw-4}, consider an amplitude with
two vector bosons and two scalars in the Abelian Higgs model in unitary
gauge,\footnote{In the 't Hooft-Feynman gauge also diagrams with
 Goldstone-boson exchange contribute, while the $WWH$ vertex vanishes in the
 massless limit. This is, however, irrelevant to the estimate of the large-$z$
 behaviour below.}
\begin{equation*}
 A_4(W_1^{-,\rh}\!\!\,H_2^{\rh},H_3^{\rh},W_4^{-,\ra}) 
=\parbox{.7\textwidth}{
\includegraphics[width=.7\textwidth]{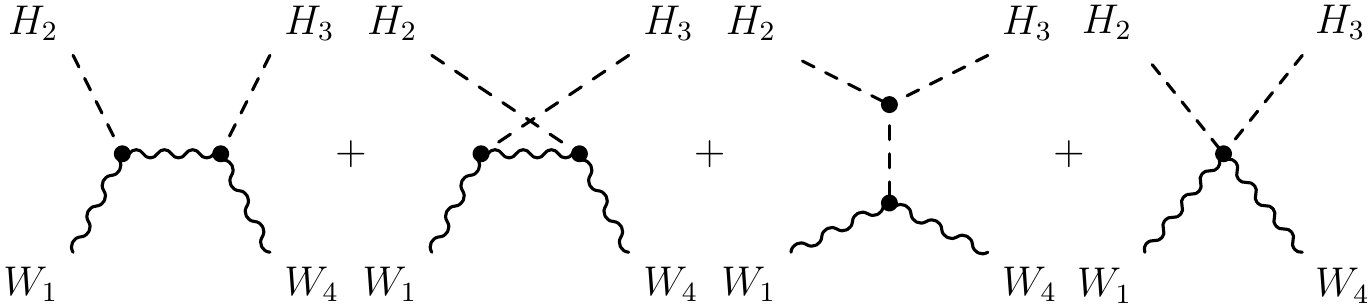}
}.
\end{equation*}
In the massless case the shift is
allowed according to the bound
$ \gamma_{\text{BCFW}}^{\rh,m=0}\leq -1$, whereas in the massive case the
weaker estimate $\gamma_{\text{BCFW}}^{\rh}\leq 0$ is obtained
from~\eqref{eq:bound-bcfw-hol}. This difference can be traced to the
different role of the reference spinors $q_i$ in the massless and massive
cases. The diagram with the four-point vertex is proportional to
\begin{equation}
 \label{eq:epsh-epsa}
  \hat
 \epsilon^{\rh}(k_1,-)\cdot\hat\epsilon^{\ra}(k_4,-)=\frac{\braket{k^\flat_1k^\flat_4}\sbraket{q_1 q_4}}{\sbraket{k^\flat_1q_1}\sbraket{k^\flat_4\eta}},
\end{equation}
which can be made to vanish in the massless case by choosing equal reference
spinors, without affecting the result for the
amplitude. In the massive case the choice of reference spinors determines the
spin axis and therefore the physical result. For the choice of reference
spinors made in Section~\ref{sec:construct}, the scalar product does not
vanish so the bound~\eqref{eq:bound-bcfw-hol} is saturated.
The same conclusion is reached using the Ward
identity~\eqref{eq:gb-et-2},
\begin{equation}
 A_4(W_1^{-,\rh}\!\!\,H_2^{\rh},H_3^{\rh},W_4^{-,\ra})\propto\frac{1}{z}
 \left( m_{W_4} A_4(W_1^{-,\rh}\!\!\,H_2^{\rh},H_3^{\rh},\phi_4^{\ra})-k_{4,\mu}A_4^\mu(W_1^{-,\rh}\!\!\,H_2^{\rh},H_3^{\rh},W_4^{\ra})\right).
\end{equation}
Focusing on the second term, the diagram with the four-point vertex is proportional to
\begin{equation}
  \hat
 \epsilon^{\rh}(k_1,-)\cdot
 k_4=\frac{\braket{k^\flat_1k^\flat_4}\sbraket{k_4^\flat q_1}
 }{\sbraket{k^\flat_1q_1}} +\frac{m_{W_4}^2}{2(q_4\cdot k_4)} \frac{\braket{\hat
   k^\flat_1(z)q_4}\sbraket{q_4q_1}}{\sbraket{k^\flat_1q_1}}\propto z,
\end{equation}
so again equal reference spinors are required to obtain a valid shift. It is
obvious that the problematic term is absent in the massless case.
As for the example in Yukawa theory, in this case a valid Risager shift is available.

\section{On-shell constructible amplitudes}
\label{sec:construct}

In this section we establish that the following shifts are sufficient to
construct all amplitudes in SBGTs with at least five legs (or six legs for
all-fermionic amplitudes):
\begin{itemize}
\item Three-line and in some cases two-line shifts are sufficient for the
 following amplitudes:
 \begin{itemize}
 \item All amplitudes with at least two transverse vector bosons.
 \item Amplitudes with scalars and at least two vector bosons, at
  least one of which is transverse.
 \item All amplitudes that contain only fermions; fermions and vector bosons;
  fermions and SM-like Higgs bosons; or generic scalars and at least two
  fermion pairs.
 \end{itemize}
\item Five-line shifts are required
 for amplitudes with only
 scalars.
\item Four-line shifts are sufficient for all other cases.
\end{itemize}
Here ``SM-like" Higgs bosons are defined as scalars that couple to vector
bosons through vertices of type $HWW$ but not through $HHW$ vertices. 
Furthermore, three-line shifts are sufficient for amplitudes of scalars
charged under an unbroken $U(1)$ symmetry~\cite{Cheung:2015cba}, which is,
however, not present for the SM Higgs boson in the broken phase. The above
conditions apply when all legs are massive. For
at least two massless vector bosons or fermions, the familiar conditions for
massless BCFW
shifts~\cite{Britto:2005fq,Schwinn:2007ee,ArkaniHamed:2008yf,Cheung:2008dn}
can be used. We do not consider the case of shifts with one massless particle
explicitly; see~\cite{Schwinn:2007ee} for a discussion of shifts with massive
and massless quarks and gluons. Compared to the results for an unbroken gauge
theory with fermionic and scalar matter fields~\cite{Cheung:2015cba}, the new
feature are four-line shifts for longitudinal gauge bosons, which arise by
exploiting little-group transformations. 

The above statements are derived in the remainder of this section. In the
study of the $z\to\infty$ behaviour we make use of the ability to perform
different types of shifts for different spin states of the external
particles. This requires a choice of spin axes for the shifted particles so
that all required shifts are feasible. Such a choice is introduced in
Section~\ref{sec:frame}. The conditions for allowed shifts for this setup are
summarized in Section~\ref{sec:shift-frame} while the minimal shifts required
for the construction of different classes of amplitudes are investigated in
Section~\ref{sec:minimal-shift}.

\subsection{Choice of reference spinors for all spin configurations}
\label{sec:frame}

To define the spin axes for all particles in a way that allows to perform
different types of shifts depending on the spins of the
particles, two ``reference particles" will be singled out. For definiteness,
these particles will be assigned the momenta $k_1$ and $k_n$. Since the choice
of a spin axis is not necessary for purely scalar amplitudes, in all relevant
cases it is possible to choose at least one particle with spin as reference
particle. As in~\eqref{eq:decompmom} the reference momenta can be expressed in
terms of two light-like vectors $l_{1/n}$ according to
\begin{align}
k_1& = l_1 + \alpha_n l_n, 
& 
k_n &= \alpha_1 l_1 + l_n.
\end{align}
The reference spinors for the two selected legs are chosen as
\begin{align}
 q_{1,\alpha}&=l_{n,\alpha}, & q_{1,\dot \alpha}&=l_{n,\dot \alpha},\label{eq:q1} \\
 q_{n,\alpha}&=l_{1,\alpha}, & q_{n,\dot \alpha}&=l_{1,\dot \alpha},\label{eq:qn}
\end{align}
while those for all other legs are taken as
\begin{align}
  q_{i,\alpha}&=l_{n,\alpha}, & q_{i,\dot \alpha}&=l_{1,\dot \alpha}.
\label{eq:qi}
\end{align}
A similar construction was used in~\cite{Schwinn:2007ee}.

Provided the amplitudes for all spin states can be computed recursively with
this choice of spin axes, the amplitudes for arbitrary spin axes follow
from little-group transformations. It is even sufficient to keep the spin
quantum number of one particle fixed, for instance by taking $s_n$ as negative. This information allows to
reconstruct the amplitudes for arbitrary spin axes of the remaining legs using
the little-group transformations~\eqref{eq:little-dirac-spin} and~\eqref{eq:little-vector-spin},
\begin{equation}
 A(\Phi_1^{s_1'},\dots,\Phi_{n-1}^{s_{n-1}'},\Phi_n^-)=\sum_{s_1}\dots\sum_{s_{n-1}}
\mathcal{R}^{(S_1)}_{s_1',s_1}\dots \mathcal{R}^{(S_{n-1})}_{s_{n-1}',s_{n-1}}A(\Phi_1^{s_1},\dots,\Phi_{n-1}^{s_{n-1}},\Phi_n^-).
\end{equation}
Since now the spin axes for all particles are independent, the operator $J_{k_n}^+$~\eqref{eq:generators-spin}
can be used to raise the spin of the last particle.

\subsection{Explicit form of shifts}
\label{sec:shift-frame}
The choice of reference spinors of Section~\ref{sec:frame} allows to perform
all of the types of shifts constructed in Section~\ref{sec:shift}.
We summarize the explicit form and the conditions for the shifted legs in
order to obtain an allowed shift.

\paragraph{Two-line BCFW-type shift} The choice of reference
spinors~\eqref{eq:q1} and ~\eqref{eq:qn} allows to
shift the two reference particles by a two-line BCFW shift~\eqref{eq:bcfw2},
\begin{align}
 \hat k_{1,\alpha}^{\flat,\rh}(z) &=l_{1,\alpha}+z l_{n,\alpha}, &
 \hat k_{1,\dot\alpha}^{\flat,\rh}(z)&=l_{1,\dot\alpha},\nonumber\\
\hat k_{n,\alpha}^{\flat,\ra}(z)&= l_{n,\alpha}, &
 \hat k_{n,\dot\alpha}^{\flat\ra}(z)&=l_{n,\dot\alpha}-z l_{1,\dot \alpha}.
\end{align}
For the internal lines in the recursion relation~\eqref{eq:recurse}, the
reference spinors
$ q_{\mathcal{F},\dot\alpha}=\eta_{\dot\alpha}=l_{1,\dot\alpha}$ and
$ q_{\mathcal{F},\alpha}=\eta_{\alpha}=l_{n,\alpha}$ are the same as for
the external legs~\eqref{eq:qi}. A valid recursion relation is obtained from
the bound~\eqref{eq:bound-generic} if
\begin{equation}
 \label{eq:bound-bcfw2}
 \gamma\leq 1 - s_1+ s_n<0.
\end{equation}
This allows to shift the following combinations of
particles,
\begin{equation}
\label{eq:bcfw-2}
 W_1^{+,\rh}W_n^{-,\ra}, \quad \psi_1^{+,\rh} W_n^{-,\ra},\quad W_1^{+,\rh}\psi_n^{-,\ra} ,\quad 
  \psi_1^{+,\rh} \psi_n^{-,\ra}.
 \end{equation}
 Two fermions may only be shifted if they belong to different fermion
 lines~\cite{Schwinn:2007ee} since in this case the skeleton amplitude
 necessarily includes two fermionic background insertions so that the
 condition~\eqref{eq:bound-bcfw2} improves to
\begin{equation}
 \gamma\leq - s_1+ s_n <0.
\end{equation}
For amplitudes with massless particles not all allowed shifts are found by
this simple power-counting analysis; in particular shifts of particles with
identical helicities are possible~\cite{Britto:2005fq}. We argue at the end of
Section~\ref{sec:minimal-shift} that similar improvements are not expected for
shifts of massive particles.

\paragraph{Holomorphic Risager-type shift}
An $h$-line shift, which may include the reference momentum $k_1$ but not
$k_n$, is possible with the shift spinor $\eta_\alpha=l_{n,\alpha}$ and
the shifted momenta
\begin{align}
 \hat k_{i,\alpha}^{\flat,\rh}(z) &=k_{i,\alpha}^{\flat}+zc_i l_{n,\alpha}, &
 \hat k_{i,\dot\alpha}^{\flat,\rh}(z)&=k^\flat_{i,\dot\alpha}.
\end{align}
Therefore $h+1$-point
functions can be constructed with an $h$-line Risager shift for the
above choice of spin axes.\footnote{Since a choice of spin axes is not necessary for all-scalar
amplitudes, the five-scalar amplitude can be constructed from a
five-line shift.}
The reference spinors for the internal line in the recursion~\eqref{eq:recurse}
are fixed according to~\eqref{eq:ref-internal-risager}, which implies that the
holomorphic reference spinor $ q_{\mathcal{F},\alpha}=\eta_\alpha=l_{n,\alpha}$ in the
subamplitudes is the same as for the full amplitude. 
According to the bound~\eqref{eq:risager-scale}, a valid $h$-line recursion
relation for dimensionless coupling constants is
obtained if the spin projections of the shifted legs satisfy
\begin{equation}
 \sum_{\mathcal{S}}s_i > 4-h.
\end{equation}

\paragraph{Anti-holomorphic Risager-type shift} In an $h$-line shift with shift spinor $\eta_{\dot \alpha}=l_{1,\dot
  \alpha}$ all legs including the reference momentum $k_n$, but not $k_1$,
 can be shifted as 
\begin{align}
 \hat k_{j,\alpha}^{\flat,\ra}(z) &=k_{j,\alpha}^{\flat}, &
 \hat k_{j,\dot\alpha}^{\flat,\ra}(z)&=k^\flat_{j,\dot\alpha} +
                       zd_j l_{1\dot \alpha} . 
\label{eq:anti-risager}
\end{align}
The subamplitudes inherit the anti-holomorphic reference spinor,
$ q_{\mathcal{F},\dot \alpha}=\eta_{\dot\alpha}=l_{1,\dot \alpha}$.
A valid recursion relation is
obtained for
\begin{equation}
\sum_{\mathcal{S}}s_j<-(4-h).
\end{equation}

\paragraph{Mostly holomorphic BCFW-type shift} 
An holomorphic shift is applied to  $h-1$ momenta (which
 may include $k_1$) while the reference momentum $k_n$ is
shifted anti-holomorphically,
\begin{align}
 \hat k_{i,\alpha}^{\flat,\rh}(z) &=k_{i,\alpha}^{\flat}+zc_i l_{n,\alpha}, &
 \hat k_{i,\dot\alpha}^{\flat,\rh}(z)&=k^\flat_{i,\dot\alpha},\nonumber\\
\hat k_{n,\alpha}^{\flat,\ra}(z)& =l_{n,\alpha}, &
 \hat k_{n,\dot\alpha}^{\flat\ra}(z)&=l_{n,\dot\alpha}+zd_n l_{1,\dot \alpha}.
\end{align}
As for the holomorphic Risager shift, the subamplitudes have the same common
holomorphic reference spinor $ q_{\mathcal{F},\alpha}=\eta_\alpha=l_{n,\alpha}$
according to~\eqref{eq:ref-internal-bcfw}.
If leg $n$ has negative spin projection,
a valid recursion relation is
obtained from~\eqref{eq:bound-bcfw-hol} for
\begin{equation}
 \label{eq:bound-bcfw}
 \sum_{\mathcal{S}}s_i-2 s_n^{\ra}=\sum_{\mathcal{H}}s_i+| s_n^{\ra}|>4-h.
\end{equation}

\paragraph{Mostly anti-holomorphic BCFW-type shift} Here the reference momentum
$k_1$ is shifted holomorphically and $h-1$ momenta (which may include $k_n$) are shifted anti-holomorphically, 
\begin{align}
  \hat k_{1,\alpha}^{\flat,\rh}(z) &=l_{1,\alpha}+zc_1 l_{n,\alpha}, &
  \hat k_{1,\dot\alpha}^{\flat,\rh}(z)&=l_{1,\dot\alpha},\\
  \hat k_{j,\alpha}^{\flat,\ra}(z)&= k^\flat_{j,\alpha}, &
  \hat k_{j,\dot\alpha}^{\flat\ra}(z)&=k^\flat_{j,\dot\alpha}+zd_j l_{1,\dot \alpha}.
\end{align}
The subamplitudes have the same common
anti-holomorphic reference spinor $ q_{\mathcal{F},\dot\alpha}=\eta_{\dot\alpha}=l_{1,\dot\alpha}$.
If leg one has positive spin projection, a valid recursion relation is
obtained for
\begin{equation}
 \label{eq:bound-anti-bcfw}
 \sum_{\mathcal{S}}s_j-2 s_1^{\rh}=\sum_{\mathcal{A}}s_j-| s_1^{\rh}|<-(4-h).
\end{equation}

\subsection{Minimal required shifts}
\label{sec:minimal-shift}
It is now possible to identify the minimal number of shifted legs necessary to
construct a given amplitude and verify the claims made in the beginning of
this section. Recall that little-group transformations allow to reconstruct
amplitudes with general spin quantum numbers from amplitudes where the reference
particle $n$ has spin $-\frac{1}{2}$ or $-1$, provided the amplitudes for arbitrary
spin configurations of the remaining particles are known for a fixed choice of
reference legs $1$ and $n$.

\subsubsection{Five-line constructible amplitudes}
All amplitudes are constructible either from holomorphic or anti-holomorphic
five-line Risager shifts since the condition $|\sum_{\mathcal{S}}s_i|\geq 0$ is
obviously always satisfied. This agrees with the conclusion from the massless case~\cite{Cheung:2015cba}.

\subsubsection{Four-line constructible amplitudes}
The following four-line shifts are possible:
\begin{itemize}
\item Holomorphic (anti-holomorphic) Risager shifts for $\sum_{\mathcal{H}}s_i> 0$ ($\sum_{\mathcal{A}}s_j< 0$).
\item Mostly holomorphic BCFW-type shifts provided the
 holomorphically shifted particles satisfy $
 \sum_{\mathcal{H}}s_i>-|s_n| $ for $s_n<0$. 
\item Mostly anti-holomorphic BCFW-type shifts provided the
 anti-holomorphically shifted particles satisfy
 $ \sum_{\mathcal{A}}s_j<s_1 $ for $s_1>0$.
\end{itemize}
Therefore all amplitudes are four-line Risager constructible unless
$\sum_{\mathcal{S}}s_i=0$. In this case, amplitudes with at least one fermion
pair or vector boson can be constructed with a BCFW-type shift with a particle
of negative spin projection as reference leg $n$. The only amplitudes that
are not four-line constructible are therefore amplitudes with only
scalars.\footnote{Amplitudes with only longitudinal vector bosons can be
 obtained from the allowed four-line BCFW shift
 $W^{0,\rh}W^{0,\rh}W^{0,\rh}W_n^{-,\ra}$ using the little group raising
 operator on the reference leg with negative spin.}

\subsubsection{Three-line constructible amplitudes}
The following three-line shifts are possible:
\begin{itemize}
\item Holomorphic (anti-holomorphic) Risager shifts for $\sum_{\mathcal{H}}s_i> 1$ ($\sum_{\mathcal{A}}s_j< -1$).
\item Mostly holomorphic BCFW-type shifts provided the
 holomorphically shifted particles satisfy $
 \sum_{\mathcal{H}}s_i>1-|s_n| $ for $s_n<0$.
\item Mostly anti-holomorphic BCFW-type shifts provided the
 anti-holomorphically shifted particles satisfy $
 \sum_{\mathcal{A}}s_j<s_1-1 $ for $s_1>0$.
\end{itemize}
We now determine all amplitudes that can be constructed using three-line shifts,
ensuring that all spin configurations can be computed
for a fixed choice of reference particles.
In some cases, also the two-line shifts~\eqref{eq:bcfw-2} can be used.

\paragraph{Vector-boson amplitudes}
Two vector bosons $W_1^{s_1}$ and $W_n^-$ are selected as reference particles.
A two-line BCFW shift is possible for the spin configuration
$W_1^{+,\rh} W_n^{-,\ra}$. For other spin projections of $W_1$,
anti-holomorphic Risager shifts $W_i^{s_i,\ra}W_j^{s_j,\ra} W_n^{-,\ra}$ can
be applied if the amplitude contains two further vector bosons with $s_i+s_j\leq -1$ while mostly holomorphic BCFW shifts
$W_i^{s_i,\rh}W_j^{s_j,\rh} W_n^{-,\ra}$ are possible for $s_i+s_j \geq 1$.
These shifts cover all amplitudes with at least five vector bosons apart from
those where all vector bosons besides the two reference legs are longitudinal,
where four-line shifts can be applied as discussed above.

\paragraph{Amplitudes with scalars and vector bosons} 
A scalar $\phi_1$ and a vector boson $W_n^-$ can be selected as reference
particles. For amplitudes with at least one additional transverse vector boson
$W_i$, either a mostly-holomorphic BCFW shift
$ \phi_1^{\rh}W_i^{+,\rh} W_n^{-,\ra}$ is possible or one can select a further
particle $\Phi_j\in\{W_j^-,W_j^0,\phi_j\}$ so that an anti-holomorphic
Risager shift $ W_i^{-,\ra}\Phi_j^{\ra} W_n^{-,\ra}$ can be performed.
This covers all amplitudes with scalars and
at least two vector bosons, unless all vector bosons besides $W_n^-$ are
longitudinal where again a four-line shift is required.

For amplitudes with scalars and only one vector boson, neither the condition
$\sum_{\mathcal{A}}s_j<-1$ for a Risager shift or $\sum_{\mathcal{H}}s_i> 0$
for a mostly holomorphic BCFW shift can be satisfied. However, such
amplitudes do not appear for SM-like Higgs bosons, which only couple to pairs
of vector bosons.

\paragraph{Amplitudes with scalars and fermions}
A scalar $\phi_1$ and a fermion
$\psi_n^-$ can be chosen as reference particles. For amplitudes with at least
two fermion pairs there must be two additional fermions with equal spin
quantum number so that a mostly-holomorphic BCFW shift $
\psi_i^{+,\rh}\psi_j^{+,\rh} \psi_n^{-,\ra}$ or an anti-holomorphic Risager
shift $ \psi_i^{-,\ra}\psi_j^{-,\ra} \psi_n^{-,\ra}$ are possible.

For amplitudes with only one fermion pair, the conditions for a three-line
shift cannot be satisfied for generic scalar particles. However, the
situation improves for SM-like Higgs bosons without $HHW$ couplings. In this
case there are no internal vector-boson lines and at least one scalar
background insertion must appear in the skeleton amplitudes for the shifts
$ \phi^{\rh} \psi^{+,\rh}\psi_n^{-,\ra}$ and
$\phi^{\ra} \psi^{-,\ra}\psi_n^{-,\ra}$. This improves the scaling, for
example in case of the BCFW-type shift~\eqref{eq:bound-bcfw-hol},
 \begin{equation}
  \gamma_{\text{BCFW}}^{\rh}  \leq
  \frac{1}{2}\left(1-b_\phi \right)- |s_n^{\ra}|
  =-\frac{1}{2},
 \end{equation}
and similarly for the Risager shift~\eqref{eq:risager-scale}.
 Therefore, all amplitudes with SM-like Higgs bosons and one fermion pair are
three-line constructible.
For the case of four-point amplitudes, this can be checked in the example given in Section~\ref{sec:bound-example}.

\paragraph{Amplitudes with a fermion pair and vector bosons}

A fermion $\psi_1^{s_1}$ and a vector boson $W_n^-$ can be taken as reference
particles. A two-line BCFW shift can be performed for the spin configuration
$\psi_1^{+,\rh} W_n^{-,\ra}$. Otherwise, a Risager shift
$\Phi^{s_i\ra}_i\Phi^{s_j,\ra}_j W_n^{-,\ra}$ or a BCFW-type shift
$\Phi^{s_i,\rh}_i\Phi^{s_j\rh}_j W_n^{-,\ra}$ can be performed if the
amplitude contains two additional particles with $s_i+s_j<0$ or $s_i+s_j>0$,
respectively. Since the amplitude must contain a second fermion with
$s=\pm\frac{1}{2}$, one of these conditions can always be fulfilled for
amplitudes with arbitrary remaining particle content.

\paragraph{Amplitudes with only fermions} 
We consider all-fermionic amplitudes with at least six legs and assume every
fermion can be uniquely assigned to one fermion line for all Feynman diagrams
contributing to the amplitude.\footnote{It is always possible to construct
 partial amplitudes with this property by assigning the fermions pairs to
 different, possibly artificial, generations and assuming flavour-mixing
 matrices to be diagonal.} Then two fermions from different fermion lines
can be chosen as reference particles $\psi_1^{s_1}$ and $\psi^{-}_n$. If
reference particle one has positive spin, a two-line BCFW shift
$\psi^{+,\rh}_1\psi^{-,\ra}_n$ is possible. For negative spin projection of
leg one, it is always possible to perform a mostly-holomorphic BCFW shift $
\psi_i^{ +,\rh}\psi_j^{+,\rh} \psi_n^{-,\ra}$ or an anti-holomorphic Risager
shift $ \psi_i^{ -,\ra}\psi_j^{-,\ra} \psi_n^{-,\ra}$ since two of the 
 remaining (at least four) fermions must have the same spin quantum number.
 This conclusion agrees
with~\cite{Schwinn:2007ee}, however, here we do not require the additional
condition that the reference particle $\psi_1$ and the three shifted legs all
belong to different fermion lines.

Therefore we have verified all cases of three-line constructible amplitudes
given in the beginning of Section~\ref{sec:construct}.

\paragraph{Further improvements?}
As mentioned above, for massless particles also two-line BCFW shifts of legs
with equal helicity are allowed, although not obvious from simple
power-counting. This may be proven for instance using an auxiliary three-line
recursion~\cite{Badger:2005zh,Schwinn:2007ee} or a background field
analysis~\cite{ArkaniHamed:2008yf}. While we have not performed a
comprehensive analysis, such improvements do not appear to be possible for
massive particles. For massive quarks, it is known that BCFW shifts of two
legs with the same spin are not allowed~\cite{Schwinn:2007ee}. As a starting
point of the inductive proof using auxiliary shifts, three-point amplitudes
involving the two shifted legs must vanish for
$z\to \infty$~\cite{Schwinn:2007ee}. For a two-line shift of two massive
vector bosons $ W_1^{-,\rh} W_n^{-,\ra}$ and the choice of reference
spinors as in Section~\ref{sec:frame}, this induction assumption is violated
since three-point functions for an unshifted vector boson $W_i^-$ and the 
two shifted vector bosons do not vanish for $z\to\infty$,
\begin{equation}
 A_3(\hat W_1^{-,\rh}, W_i^-,\hat W_n^{-,\ra})
 \propto
  \bigl(\hat\epsilon^{\rh}(k_1,-)\cdot\hat\epsilon^{\ra}(k_n,-)\bigr)\,
 \bigl( \epsilon(k_i,-)\cdot k^{\ra}_n(z)\bigr)
\sim z^0,
\end{equation}
since $ \hat \epsilon^{\rh}(k_1,-)\cdot\hat\epsilon^{\ra}(k_n,-) \neq 0$ as
in~\eqref{eq:epsh-epsa}. This is another example for the dependence of the
large-$z$ behaviour on the choice of spin axis discussed in
Section~\ref{sec:bound-example}.

\section{Application to selection rules}
\label{sec:select}

Selection rules for helicity amplitudes of massless particles with arbitrary
multiplicities are well known, see e.g.~\cite{Mangano:1990by,Dixon:1996wi} for
reviews. In particular, gluonic amplitudes vanish if all gluons have the same
helicity or only one has a different helicity. The first nonvanishing
amplitudes are the so-called maximally helicity violating~(MHV) amplitudes,
which have a very simple all-multiplicity expression. The corresponding
selection rules for amplitudes with massive particles are less restrictive and
depend on the choice of spin axes. Selection rules for all-multiplicity
amplitudes with massive vector bosons have been derived from supersymmetric
identities~\cite{Schwinn:2006ca,Boels:2011zz} or using diagrammatic
arguments~\cite{Coradeschi:2012iu}. Here we provide an alternative derivation
of some of these results as an application of on-shell recursion relations for
amplitudes with massive vector bosons. We furthermore discuss the role played
by the choice of spin axes.

For the derivation of the selection rules, it is useful to deviate slightly
from the choice of spin axes of Section~\ref{sec:frame} and choose the
reference spinors of all remaining particles in terms of the momentum and spin
axis of one reference particle $n$:
\begin{align}
 \label{eq:select-frame}
 q_{i,\alpha}&=k^\flat_{n,\alpha},& q_{i,\dot\alpha}&=q_{n,\dot\alpha}.
\end{align}
Relations with an exchanged role of positive and negative spin projections can
be derived for reference spinors $q_{i,\alpha}=q_{n,\alpha}$ and
$q_{i,\dot\alpha}=k_{n,\dot\alpha}^\flat$. On-shell recursion relations can
be used to show that amplitudes with only vector bosons with negative spin
projection vanish for the choice~\eqref{eq:select-frame}, as well as those
with one additional scalar or longitudinal vector boson,
\begin{equation}
 \label{eq:select-w-1}
 \begin{aligned}
 A(W_1^-,W_2^-,\dots, W_n^-)&=0,\\
 A(W_1^0,W_2^-,\dots, W_n^-)&=0,\\
 A(\phi_1,W_2^-,\dots, W_n^-)&=0.
 \end{aligned}
\end{equation}
Furthermore, amplitudes where the reference particle has negative spin
projection vanish if all other vector bosons have
positive spin projection, or contain one scalar or longitudinal vector boson,
\begin{equation}
 \label{eq:select-w-2}
 \begin{aligned}
 A(W_1^+,W_2^+,\dots, W_n^-)&=0,\\
 A(W_1^0,W_2^+,\dots, W_n^-)&=0,\\
 A(\phi_1,W_2^+,\dots, W_n^-)&=0.
\end{aligned}
\end{equation}
These identities are derived in Section~\ref{sec:select-boson}. In
Section~\ref{sec:select-fermion}, similar identities are derived for amplitudes
with a fermion pair and vector bosons where all spin projections are negative,
\begin{equation}
 \label{eq:select-psi-1}
 A(\bar \psi_1^-,\psi_2^-,W_3^-,\dots, W_n^-)=0,
\end{equation}
or all positive with the exception of the reference particle,
\begin{align}
   A(\bar\psi_1^+,\psi_2^+,W_3^+,\dots,\dots W_n^-)&=0, \label{eq:select-psi-2}\\
 A(\bar\psi_1^+,W_2^+,\dots \psi_n^-)&=0. \label{eq:select-psi-3}
\end{align}

All amplitudes appearing in these selection rules vanish in the massless
limit. For the massive case, however, the selection rules are valid only for
the spin axes defined by the choice of reference
spinors~\eqref{eq:select-frame}. These results are compatible with relations
valid in massive supersymmetric theories for the same choice
$q_{i,\alpha}=k_{n,\alpha}^\flat$ of holomorphic reference
spinors~\cite{Boels:2011zz}. In section~\ref{sec:select-little}, little-group
transformations are used to transform these selection rules into a frame with
a common spin axis for all particles, as used in~\cite{Coradeschi:2012iu}. In
this frame, the selection rules~\eqref{eq:select-w-1}
and~\eqref{eq:select-psi-1} continue to hold exactly, while the amplitudes
appearing in the other selection rules are mass-suppressed in the high-energy
limit.

\subsection{Selection rules for bosonic amplitudes}
\label{sec:select-boson}
For the set of amplitudes~\eqref{eq:select-w-1} it is possible to perform a
three-line anti-holomorphic Risager shift
$ W_l^{-,\ra}W_{m}^{-,\ra}W_n^{-,\ra}$ with arbitrary shift spinor
$\eta_{\dot\alpha}$ so that the amplitudes satisfy the recursion
 \begin{multline}
  \label{eq:risager-select}
 A(\Phi_1,\dots W_l^-,\dots,W_{m}^-,\dots ,W_n^-)\\
 =
 \sum_{\mathcal{F},s}A_{\mathcal{F}}(\Phi_1,\dots \hat W_l^{-,\ra}\dots\hat{\Phi}_{\mathcal{F}}^s) 
\frac{\ii}{K_{\mathcal{F}}^2 -M_{\mathcal{F}}^2} 
A_{\mathcal{F'}}(\hat {\Phi}_{\mathcal{F'}}^{-s},\dots, \hat W_m^{-,\ra}\dots \hat W_n^{-,\ra}\dots)+\dots,
\end{multline}
where $\Phi_1\in\{W^-,W^0,\phi\}$ and $\Phi_{\mathcal{F}}^s\in\{W^s,\phi\}$.
The dots indicate similar contributions with a different distribution of the
shifted legs over the factorized subamplitudes.

For the anti-holomorphic Risager shift, all
scalar products of polarization vectors with negative spin projection vanish
for all external and internal lines,
\begin{equation}
 \label{eq:select-eps--}
 \epsilon(k_i,-)\cdot\epsilon(k_j,-)=0,
\end{equation}
since all external and internal lines are defined using the same
anti-holomorphic reference spinor,
$q_{\mathcal{F},\dot\alpha}=q_{i,\dot\alpha}=q_{n,\dot\alpha}$. 
This implies the
 selection rules for three-point vertices,
\begin{equation}
 A_3(\Phi_i,W_j^-,W_k^-)=0, \label{eq:select-phiw--}
\end{equation}
where all particles may be shifted or unshifted, as well as external or
internal. The result for $\Phi_i=\{W_i^-,\phi_i\}$ follows trivially since
all terms in the vertex involve a scalar product of two polarization vectors.
For the longitudinal polarization $\Phi_i=W_i^0$, the selection rule follows
from the Ward identity~\eqref{eq:gb-wi} since the choice of reference
spinors~\eqref{eq:select-frame} implies the identity
\begin{equation}
 \epsilon(k_j,-)\cdot r_i =0
\end{equation}
for all internal or external legs, where the remainder vector $r_\mu\propto q_\mu$ is defined in~\eqref{eq:def-r}.

The recursion relation~\eqref{eq:risager-select} allows to construct the four-point amplitudes from three-point
amplitudes as input. The selection rules for the three point functions imply
that all possible combinations of three-point building blocks appearing in the
numerator for the different factorization channels vanish, i.e.
\begin{equation}
 \sum_{s} A_3(\Phi_1,\hat W_l^{-,\ra},\hat\Phi_{\mathcal{F}}^s)
 A_3(\hat\Phi_{\mathcal{F}'}^{-s},\hat W_m^{-,\ra},\hat W_{n}^{-,\ra})
  =0.
\end{equation}
This verifies the selection rules~\eqref{eq:select-w-1} for four-point
functions. By induction, the argument generalizes to general multiplicities
due to the structure of the recursion relation~\eqref{eq:risager-select}.

For the amplitudes~\eqref{eq:select-w-2} it is possible to perform a
mostly-holomorphic BCFW three-line shift~\eqref{eq:bcfw-h}
$ W_l^{+,\rh}W_{m}^{+,\rh}W_n^{-,\ra}$, which leads to a recursion relation of
a similar form as~\eqref{eq:risager-select} up to the different spin labels.
Internal lines in the recursion relations are defined by the same holomorphic
reference spinor $q_{\mathcal{F},\alpha}=q_{i,\alpha}=k_{n,\alpha}^\flat$ as
all external line apart from the reference leg $n$. Therefore the
polarization vectors with positive spin projection share the same reference
spinor for internal and external lines. This choice implies that the scalar
products
 \begin{align}
 \epsilon(k_i,+)\cdot\epsilon(k_j,+)&=0,&
 \epsilon(k_i,\pm)\cdot\hat\epsilon^{\ra}(k_n,-)&=0,
\end{align}
vanish 
where $i$ and $j$ denote arbitrary vector bosons, including
the reference leg $n$. This implies 
the selection rules 
\begin{equation}
 A_3(\Phi_i,W_j^+,W_k^+)=0
\end{equation}
with $\Phi_i=\{W_i^+,W_i^0,\phi_i\}$
for three-point functions of generic vector bosons and
\begin{equation}
A_3(\Phi_i,W_j^+,\hat W_n^{-,\ra})=0
\end{equation}
for three-point functions involving the reference leg.
The selection rules for longitudinal vector bosons hold since the choice of spin axes ensures
\begin{equation}
 \epsilon(k_j,+)\cdot r_i =0= \hat\epsilon^{\ra}(k_n,-)\cdot r_i
\end{equation}
for all legs $i$ and $j$. 
In analogy to the discussion for the Risager shift, it is seen that all
possible combinations of three-point amplitudes contributing to the recursive
construction of the four-point amplitude
 vanish,
\begin{align}
 \sum_{s} A_3(\Phi_1,\hat W_l^{+,\rh},\hat\Phi_{\mathcal{F}}^s)
 A_3(\hat\Phi_{\mathcal{F}'}^{-s},\hat W_m^{+,\rh},\hat W_{n}^{-,\ra})
  &=0,\\
\sum_{s} A_3(\hat W_l^{+,\rh},\hat W_m^{+,\rh},\hat\Phi_{\mathcal{F}}^s)
 A_3(\hat\Phi_{\mathcal{F}'}^{-s},\Phi_1,\hat W_{n}^{-,\ra})
  &=0.
 \end{align}
The selection rules~\eqref{eq:select-w-2} for arbitrary multiplicities
 follow by induction using the recursion relation.
\subsection{Selection rules for amplitudes with a fermion pair}
\label{sec:select-fermion}

Similar to the bosonic case, the selection rule~\eqref{eq:select-psi-1} for a
fermion pair and an arbitrary number of vector bosons with identical spin
labels can be established by a three-line anti-holomorphic Risager shift
$ \psi_l^{-,\ra}W_{m}^{-,\ra}W_n^{-,\ra}$. In the recursion relation,
contributions from internal boson and fermion lines must be taken into account,
\begin{align}
 \label{eq:recurse-psi-1}
 & A(\bar\psi_1^-,\dots,\psi_l^-,\dots,W_{m}^-,\dots ,W_n^-)\nonumber\\
 & =
 \sum_{\mathcal{F},s}A_{\mathcal{F}}( \bar \psi_1^-,\dots\hat\psi_l^{-,\ra}\dots\hat{\Phi}_{\mathcal{F}}^s) 
\frac{\ii}{K_{\mathcal{F}}^2 -M_{\mathcal{F}}^2} 
A_{\mathcal{F'}}(\hat {\Phi}_{\mathcal{F}'}^{-s},\dots, \hat W_m^{-,\ra},\dots \hat
W_n^{-,\ra}\dots)\nonumber\\
&+ \sum_{\mathcal{F},s}A_{\mathcal{F}}( \bar \psi_1^-,\dots\hat W_m^{-,\ra}\dots\hat{\psi}_{\mathcal{F}}^s) 
\frac{\ii}{K_{\mathcal{F}}^2 -M_{\mathcal{F}}^2} 
A_{\mathcal{F'}}(\hat {\bar\psi}_{\mathcal{F}'}^{-s},\dots, \hat\psi_l^{-,\ra},\dots \hat
W_n^{-,\ra}\dots)+\dots,
\end{align}
where again the dots indicate contributions with a different distribution of the shifted legs
over the subamplitudes.

Selection rules for the three-point fermion amplitudes can be inferred from
the contractions of Dirac spinors~\eqref{eq:spinors} with the polarization
vectors~\eqref{eq:pol},
\begin{equation}
 \fmslash\epsilon(k_i,-)u(k_j,-\tfrac{1}{2})\propto
 \begin{pmatrix}
   - k_{i,\alpha}^\flat\,m_j\frac{ \sbraket{q_iq_j}}{\sbraket{q_jk_j^\flat}} \\
   q_{i,\dot\alpha}\braket{k_i^\flat k_j^\flat}
   \end{pmatrix}.
\end{equation}
This implies the selection rule for three-point functions for generic legs,\footnote{For simplicity vector-like couplings are
 assumed. The presence of different left- and right-handed couplings does not
 change the results, while for purely chiral couplings additional selection
 rules can arise.}
\begin{equation}
 \label{eq:select-psi-3pt}
  A_3(\psi_i^-,W_j^-,\psi_k^-)=0,
\end{equation}
since the same anti-holomorphic reference spinors
 enter the left-handed
polarization vectors and spinors of all external or internal particles,
including the reference leg.

In the recursive construction of the four-point amplitude
using~\eqref{eq:recurse-psi-1}, the contributing products of three-point
amplitudes all vanish for arbitrary spin states of the intermediate particle,
\begin{align}
  \sum_{s} A_3(\bar\psi_1^-,\hat\psi_l^{-,\ra},\hat\Phi_{\mathcal{F}}^s)
 A_3(\hat\Phi_{\mathcal{F}'}^{-s},\hat W_m^{-,\ra},\hat W_{n}^{-,\ra})
 &=0,\\
  \sum_{s}
 A_3(\bar\psi_1^-,\hat W_{m}^{-,\ra},\hat\psi_{\mathcal{F}}^s)
 A_3(\hat{\bar\psi}_{\mathcal{F}'}^{-s},\hat\psi_{l}^{-,\ra},\hat W_{n}^{-,\ra})
 &=0.
\end{align}
Note that the selection rules for the bosonic
vertices~\eqref{eq:select-phiw--} ensure that the properties of the vertices
involving fermions and scalars or longitudinal vector bosons are not required
to show that the four-point amplitude vanishes. Due to the structure of the
recursion relation~\eqref{eq:recurse-psi-1}, the selection
rule~\eqref{eq:select-psi-1} follows from induction.

The selection rules~\eqref{eq:select-psi-2} and~\eqref{eq:select-psi-3} can be
derived using a mostly-holomorphic BCFW-type shift
$\psi_l^{+,\rh}W_{m}^{+,\rh}W_n^{-,\ra}$ or
$W_l^{+,\rh}W_{m}^{+,\rh}\psi_n^{-,\ra}$.
In analogy to~\eqref{eq:select-psi-3pt}, the identities
\begin{align}
  A_3(\psi_i^+,W_j^+,\psi_k^+)&=0 
\end{align}
 hold since the same holomorphic
reference spinor enters all wave functions of particles with positive spin
projection. For three-point amplitudes involving the reference particles,
selection rules can be inferred from the contractions
\begin{align} 
  \fmslash\epsilon(k_i,+)\hat u^{\ra}(k_n,-\tfrac{1}{2})&\propto
 \begin{pmatrix}
   -q_{i,\alpha}\,m_n\frac{ \sbraket{k^\flat_iq_n}}{\sbraket{q_nk_n^\flat}} \\
    k^\flat_{i,\dot\alpha}\braket{q_i k_n^\flat}
  \end{pmatrix},&
 \fmslash{\hat \epsilon}^{\ra}(k_n,-)u(k_j,+\tfrac{1}{2})&\propto
 \begin{pmatrix}
    k^\flat_{n,\alpha}\sbraket{q_n k_j^\flat}\\
    -q_{n,\dot\alpha}\,m_j\frac{ \braket{k^\flat_nq_j}}{\braket{k_j^\flat q_j}} \\
   \end{pmatrix}.            
\end{align}
For the above choice of reference spinors, these imply the selection rules
\begin{align}
  A_3(\bar\psi_i^+,W_j^+,\hat\psi_n^{-,\ra})&=0,&  A_3(\bar\psi_i^+,\psi_k^+,\hat W_n^{-,\ra})&=0 ,
\end{align}
if a vector boson or fermion is used as reference particle.

The selection rule~\eqref{eq:select-psi-2} is again derived inductively from
the recursion relation, which has a similar form
as~\eqref{eq:recurse-psi-1}. For the four-point amplitude this is seen since
all products of three-point functions appearing in the recursive construction
vanish,
\begin{align}
  \sum_{\Phi,s} A_3(\bar\psi_1^+,\hat\psi_l^{+,\rh},\hat\Phi_{\mathcal{F}}^s)
 A_3(\hat\Phi_{\mathcal{F}'}^{-s},\hat W_m^{+,\rh},\hat W_{n}^{-,\ra})
 &=0,\\
  \sum_{s}
 A_3(\bar\psi_1^+,\hat W_{m}^{+,\rh},\hat\psi_{\mathcal{F}}^s)
 A_3(\hat{\bar\psi}_{\mathcal{F}'}^{-s},\hat\psi_{l}^{+,\rh},\hat W_{n}^{-,\ra})
 &=0.
\end{align}
The selection rule~\eqref{eq:select-psi-3} is derived analogously, where the
contributions 
in the four-point case are
\begin{align}
 \sum_{\Phi,s} A_3(\hat W_l^{+,\rh},\hat W_m^{+,\rh},\hat\Phi_{\mathcal{F}}^s)
 A_3(\hat\Phi_{\mathcal{F}'}^{-s},\bar\psi_1^+,\hat\psi_{n}^{-,\ra})
 &=0,\\
 \sum_s A_3(\bar\psi_1^{+},\hat W_m^{+,\rh},\hat\psi_{\mathcal{F}}^s)
 A_3(\hat{\bar\psi}_{\mathcal{F}'}^{-s},\hat W_l^{+,\rh},\hat\psi_{n}^{-,\ra})
 &=0. 
\end{align}
\subsection{Transformation to a common spin axis}
\label{sec:select-little}
 
The selection rules derived so far hold for a different spin
axis~\eqref{eq:select-frame} for legs $i\in\{1,\dots,n-1\}$ compared to leg
$n$. Little-group transformations can be used to transform to a
convention where the same reference spinors are used for all particles. Taking
the common reference spinors to be those of the reference particle
$n$, the new reference spinors are given by
\begin{align}
 q'_{i,\alpha}&=q_{n,\alpha},& q'_{i,\dot\alpha}=q_{i,\dot\alpha}=q_{n,\dot\alpha} .
\end{align}
The expressions for the little-group rotation~\eqref{eq:little-trafo} imply
that the elements of the transformation matrices for legs $i$ satisfy
$R_{+-}=0$ and $R_{--}=R_{++}^{-1}$. For the comparison with the selection
rules for a common spin axis~\cite{Coradeschi:2012iu}, the explicit expressions
of the transformations are not required, but rather the behaviour in the
high-energy limit $E\gg m_W,m_\psi$. Using the scaling
$\ket{k_i}\sim \sket{k_i}\sim \sqrt{E_i}$, the matrix elements behave in the
high-energy limit as\footnote{These relations hold independent of the scaling
 of the reference spinors in the high-energy limit.}
 \begin{align}
\label{eq:scale-trafo}  
  R_{--}&\sim 1, & R_{-+}&\sim \frac{m_i}{E_i}, & R_{++}&\sim 1.
 \end{align}
 Therefore the matrices implementing the transformation of the Dirac
 spinors~\eqref{eq:little-dirac} and vector-boson polarization
 vectors~\eqref{eq:little-vector} are of upper triangular form, with the
 off-diagonal elements suppressed by powers of $m/E$,
\begin{align}
 \begin{pmatrix}
  u'(+\tfrac{1}{2})\\[.1cm]
  u'(-\tfrac{1}{2}) \end{pmatrix}
 &=
 \begin{pmatrix}
   \mathcal{O}(1) & \mathcal{O}(\tfrac{m_\psi}{E}) \\[.1cm]
   0 & \mathcal{O}(1) & 
  \end{pmatrix}
 \begin{pmatrix}
  u(+\tfrac{1}{2})\\[.1cm]
  u(-\tfrac{1}{2})
 \end{pmatrix} ,               
\end{align}
and
\begin{equation}
 \begin{pmatrix}
   \epsilon'(+)\\ \epsilon'(0)\\ \epsilon'(-)
 \end{pmatrix}
 =  \begin{pmatrix}
   \mathcal{O}(1) & \mathcal{O}(\tfrac{m_W}{E}) & \mathcal{O}((\tfrac{m_W}{E})^2)\\
   0 & 1 & \mathcal{O}(\tfrac{m_W}{E}) \\
   0 & 0 & \mathcal{O}(1)
  \end{pmatrix}
 \begin{pmatrix}
   \epsilon(+)\\ \epsilon(0)\\ \epsilon(-)
 \end{pmatrix}  .
\end{equation}

The resulting expressions for $ \epsilon'(-)$, $ \epsilon'(0)$ and $u'(-\tfrac{1}{2})$
imply that the selection rules~\eqref{eq:select-w-1}
and~\eqref{eq:select-psi-1} for equal spin quantum numbers also hold exactly
for a common spin axis. In contrast, the
amplitudes~\eqref{eq:select-w-2},~\eqref{eq:select-psi-2}, 
and~\eqref{eq:select-psi-3} do not vanish for a common spin axis but are
suppressed by powers of $m/E$. For the amplitudes with only vector bosons, one finds
\begin{align}
  A'(W_1^+,\dots,W_n^-)&\sim \sum_{i,j}\left(\frac{m_{W_i}m_{W_j}}{E_iE_j}\right)
       A(W_1^+,\dots W_i^0,\dots,W_j^0,\dots,W_n^-) &\nonumber\\
&\quad+ \sum_i\left(\frac{m_{W_i}}{E_i}\right)^2
             A(W_1^+,\dots W_i^-,\dots,W_n^-)+\dots ,\\
  A'(W_1^0,W_2^+,\dots,W_n^-)&\sim \left(\frac{m_{W_1}}{E_1}\right)
                A(W_1^-,W_2^+,\dots,W_n^-) \nonumber\\
 &\quad + \sum_i\left(\frac{m_{W_i}}{E_i}\right)
                A(W_1^0,W_2^+,\dots W_i^0,\dots,W_n^-)\dots,
\end{align}
where the dots indicate contributions that are suppressed by higher powers of
$(m/E)$. The amplitudes with two negative spin labels on the right-hand side
of these identities are non-vanishing in the massless limit, where they are
just the MHV amplitudes. The same holds for the amplitudes
$A(\phi_1,W_2^+,\dots \phi_i,\dots,W_n^-)$, which arise from the high-energy
limit of amplitudes with two longitudinal vector bosons. Therefore, all
suppression factors are explicit in the above relations. These results agree
with Table 1 in~\cite{Coradeschi:2012iu}.

For amplitudes with a scalar one obtains
\begin{equation}
  A'(\phi_1,W_2^+,\dots,W_n^-)\sim \sum_i\left(\frac{m_{W_i}}{E_i}\right)
  A(\phi_1,W_2^+,\dots W_i^0,\dots,W_n^-)+\dots
 \end{equation}
In the high-energy limit, the amplitude on the right-hand side again tend to
MHV-type amplitudes with two scalars. 

For the fermionic amplitudes~\eqref{eq:select-psi-2} the little-group
transformation generates two types of contributions with a $(m/E)^1$
suppression factor,
\begin{align}                 
  A'(\bar\psi_1^+,\psi_2^+,W_3^+,\dots W_n^-)&\sim  \sum_{i=1,2}\left(\frac{m_{\psi_i}}{E_i}\right)A(\bar\psi_1^+,\psi_i^-,W_3^+,\dots,W_n^-)\nonumber\\
&\quad+
                                                                               \sum_{i}\left(\frac{m_{W_i}}{E_i}\right)A(\bar\psi_1^+,\psi_2^+,W_3^+,\dots,W_i^0,\dots W_n^-) +\dots
\end{align}
Here the amplitudes in the first term on the right-hand side are of MHV type
and therefore un-suppressed in the high-energy limit. Also the amplitudes in
the second line are non-vanishing in the high-energy limit, as can be seen by
applying the Goldstone-boson equivalence theorem and using the helicity
structure of the Yukawa coupling.\footnote{Here the high-energy limit is
 interpreted as the limit $(v/E)\to 0$ with the scalar vacuum expectation
 value $v$. If the limit $m_\psi\to 0$ is instead implemented by sending the
 fermion Yukawa couplings to zero, these amplitudes vanish.} Transforming the
selection rule~\eqref{eq:select-psi-3} to a common spin
axis yields 
\begin{align}     
 A'(\bar\psi_1^+,W_2^+,\dots \psi_n^-)&\sim
\left(\frac{m_{\psi_1}}{E_1}\right)A(\bar\psi_1^-,W_2^+,\dots\psi_n^-)\nonumber\\
 &\quad+\sum_i\left(\frac{m_{W_i}}{E_i}\right)A(\bar\psi_1^+,W_2^+,\dots
                     W_i^0,\dots\psi_n^-)\nonumber\\
  &\quad+\sum_i\left(\frac{m_{W_i}}{E_i}\right)^2A(\bar\psi_1^+,W_2^+,\dots W_i^-,\dots\psi_n^-)+\dots
\end{align}
Those amplitudes on the right-hand side that are multiplied by a factor of
$(m/E)^1$ are not of MHV type and vanish in the massless limit by helicity
conservation, in case of the amplitudes involving longitudinal vector bosons
after application of the Goldstone-boson equivalence theorem. Therefore these
amplitudes are themselves suppressed by a factor of $m_\psi/E$ relative to the
MHV-type amplitudes in the last line, so that all terms on the right-hand side
are effectively of order $\sim m_im_j/E^2$ for different combinations of
fermion or vector-boson masses. This is consistent with all-multiplicity
results for amplitudes for massive quarks and massless vector
bosons~\cite{Ferrario:2006np,Schwinn:2006ca}.
 
The above observations shed some new light on the results
of~\cite{Coradeschi:2012iu}: amplitudes that are mass-suppressed for common
spin axes may vanish exactly for some other choice of spin axes. In fact, the
scaling~\eqref{eq:scale-trafo} is not limited to the particular
choice~\eqref{eq:select-frame} of the initial spin axes but holds generically
in the high-energy limit. However, in general $R_{+-}$ is non-vanishing but
of order $m/E$, so the sets of amplitudes~\eqref{eq:select-w-1}
and~\eqref{eq:select-psi-1} are expected to be mass-suppressed for general
choices of the spin axes. Therefore, amplitudes that vanish for one choice of
spin axes are mass-suppressed for generic choices. These results are
consistent with the vanishing of these amplitudes in the massless limit, but
it is interesting to see them emerge from the little-group transformations,
which also allow to obtain the power of the suppression factors.
\section{Summary and conclusions}
We have performed a comprehensive study of complex deformations of Born
amplitudes in spontaneously broken gauge theories and analysed the behaviour
for large values of the deformation parameter $z$, building on previous
studies~\cite{Schwinn:2007ee,Cohen:2010mi,Cheung:2015cba}. Based on these
results we have identified the minimal shifts necessary to obtain valid
on-shell recursion relations for amplitudes with a given particle content and
spin quantum numbers. Spontaneously broken gauge invariance has been shown to
improve the large-$z$ behaviour through the use of Ward-identities.
Since an on-shell construction of three-point and four-point amplitudes can be
performed using little-group invariance and factorization
arguments~\cite{Arkani-Hamed:2017jhn}, we have focused on amplitudes with five
or more legs. We find that two-line or three-line shifts are sufficient for
all such amplitudes involving at least two transverse gauge bosons, amplitudes
with fermions and vector bosons, and for purely fermionic amplitudes with six
or more legs. Furthermore, all amplitudes with multiple SM-like Higgs bosons
and fermions or at least one transverse vector boson are three-line
constructible. For all remaining amplitudes four-line shifts are sufficient,
with the exception of pure scalar amplitudes where five-line shifts are
required. As application, we have shown how selection rules for massive
multi-boson amplitudes follow from on-shell recursion relations and have
explored the role of the choice of spin axes using little-group transformations.

Our results for the minimal shifts are in overall agreement with those for
unbroken gauge theories with the same matter content~\cite{Cheung:2015cba}, up
to the results for longitudinal vector bosons, which only arise in the massive
case. This is intuitively expected from the analogy of the large-$z$
behaviour and the high-energy limit. However, as demonstrated by some
examples, amplitudes with massive fermions and vector bosons may show worse
large-$z$ behaviour under extended BCFW-type shifts than massless amplitudes due
to the dependence on the spin axis or the appearance of contributions that are
forbidden in the massless case by the helicity structure.

We have presented little-group covariant expressions for shifted momenta and
wave functions but made a particular choice of spin axes to find valid shifts
for all spin configurations, so our final prescriptions for the shifts are not
manifestly little-group invariant. The possibility to use the covariant form
for mixed shifts of massive and massless particles~\cite{Aoude:2019tzn}
deserves further study. The question remains if a ``spin blind" recursive
construction is possible, which does not require to specify the spin axes of
massive particles. It would be interesting to explore a possible extension of a
definition of shifts for generic polarizations of massless vector
bosons~\cite{Rodina:2016mbk} to the massive case.

While we have limited the discussion to renormalizable gauge
theories with spin $s\leq 1$, the methods used for the estimate of the
large-$z$ behaviour of massive amplitudes could be extended towards effective
field theories or to higher-spin particles. In the former case, on-shell
recursion relations for non-linearly realised effective theories of massless
particles have been studied~\cite{Cheung:2015ota} while the application to the
extension of the Standard Model by higher-dimensional operators has been
initiated recently~\cite{Aoude:2019tzn}. An initial application of on-shell
recursion relations to amplitudes with massive spin-two particles was given
in~\cite{Moynihan:2017tva}.

\section*{Acknowledgments}
The work of RF was supported by the DFG project ``Precise predictions for vector-boson
scattering at hadron colliders'' (project no. 322921231) while CS acknowledges support by the 
Heisenberg Programme of the DFG.

\appendix

\section{Conventions}
\label{app:spinors}

The spinor conventions used in this paper follow~\cite{Schwinn:2007ee}.
 The sigma matrices are defined as
$\sigma_{\alpha\dot{\beta}}^{\mu} = \left( 1 , - \vec{\sigma}
\right)$, $\bar{\sigma}^{\mu \dot{\alpha} \beta} = \left( 1 ,
 \vec{\sigma} \right)$, where $\vec{\sigma} = \left( \sigma_x,
 \sigma_y, \sigma_z \right)$ are the Pauli matrices. Four-vectors
$x^\mu$ are mapped to bi-spinors according to
\begin{align}
x_{\alpha\dot{\alpha}}&=x_\mu \sigma_{\alpha\dot{\alpha}}^\mu ,&
x^{\dot{\alpha} \alpha}&=x_\mu \bar\sigma^{\mu;\dot{\alpha} \alpha}.
\end{align}
The conventions for the two-dimensional antisymmetric tensor are given by
\begin{equation}
\varepsilon^{\alpha\beta} = \varepsilon^{\dot{\alpha}\dot{\beta}} =
\varepsilon_{\alpha\beta} = \varepsilon_{\dot{\alpha}\dot{\beta}} =
\begin{pmatrix}
 0 & 1\\ -1 & 0 \\
\end{pmatrix}.
\end{equation}
Indices of two-component Weyl spinors are raised and lowered as follows:
\begin{equation}
 k^\alpha = \varepsilon^{\alpha\beta} k_\beta,
 \;\;\;
 k^{\dot{\alpha}} = \varepsilon^{\dot{\alpha}\dot{\beta}}k_{\dot{\beta}},
 \;\;\;
 k_{\dot{\beta}} =
 k^{\dot{\alpha}} \varepsilon_{\dot{\alpha}\dot{\beta}},
 \;\;\;
 k_\beta = k^\alpha \varepsilon_{\alpha\beta}.
\label{eq:indices}
\end{equation}
In the bra-ket notation spinor products are denoted as
\begin{align}
 \braket{p q} &= p^\alpha q_\alpha, &
 \sbraket{q p} &= q_{\dot{\alpha}} p^{\dot{\alpha}},
\end{align}
while contractions of spinors with ``slashed" momenta are given by
\begin{align}
 \bra{p}\fmslash k\sket{q}& = p^\alpha k_{\alpha\dot\beta}q^{\dot\alpha},&
 \sbra{p}\fmslash k\ket{q}& = p_{\dot\alpha} k^{\dot\alpha\beta}q_{\beta}.
\end{align}
Explicit solutions for the Weyl spinors associated to a light-like momentum can be written as
\begin{align}
 k_\alpha &= \frac{e^{ \frac{\ii}{2}(\chi-\phi)}}{\sqrt{k_0+k_3}}
       \begin{pmatrix}
        -k_1+\ii k_2 \\ k_0+k_3 
       \end{pmatrix}, &
       k^{\dot\alpha} &= \frac{e^{- \frac{\ii}{2}(\chi+\phi)}}{\sqrt{k_0+k_3}}
       \begin{pmatrix}
        k_0+k_3 \\ k_1+\ii k_2        
       \end{pmatrix} ,
\end{align}
where $\phi$ is defined by $ k_0+k_3 = \left| k_0+k_3 \right| e^{\ii\phi}$.
In the convention of~\cite{Schwinn:2007ee}, the arbitrary phase $\chi$ is set
to zero. With this convention, the complex conjugate spinors for real momenta
satisfy
\begin{align}
 (k_\alpha)^*&=\text{sgn}(k^+)k_{\dot\alpha},& (k^{\dot\alpha})^*&=\text{sgn}(k^+)k^{\alpha},
\end{align}
with $k^+=k^0+k^3$. This implies the relations for the spinor products
\begin{align}
\braket{kq}^*&=\sbraket{qk}, & \sbraket{qk}^*=\braket{kq}
\end{align}
if $k^+,q^+>0$. 
The relations for the Weyl spinors for crossed momenta, $p=-k$ are
given by
\begin{align}
 p_{\alpha}&=\ii\, \text{sgn} (k^+) k_{\alpha}, & p^{\dot\alpha}&=\ii\, \text{sgn} (k^+) k^{\dot \alpha},&
 p^{\alpha}&=\ii\, \text{sgn}(k^+) k^{\alpha}, & p_{\dot\alpha}&=\ii \, \text{sgn}(k^+) k_{\dot \alpha}.
\end{align}

\section{Little-group transformations}
\label{app:little}
Since we have shown the validity of on-shell recursion relations for a
particular choice of the spin axes made in Section~\ref{sec:frame}, it is
necessary to perform little-group rotations to obtain results in general
frames. According to the dictionary~\eqref{eq:ref-spinors}, a transformation
of reference spinors
$\{q_{\alpha}, q_{\dot\alpha}\}\to \{q'_{\alpha}, q'_{\dot\alpha}\}$ can be
obtained by finding a little-group transformation that takes
$k^2_\alpha\to k^{'2}_\alpha$ and $k_{\dot\alpha,2}\to k^{'}_{\dot \alpha,2}$
(i.e. $k^1_{\dot \alpha}\to k^{'1}_{\dot\alpha}$). This is achieved by the
transformation with components
 \begin{align}
 R^1{}_I&=\frac{1}{m}\sbraket{k^{'1}k_I},&
 R^2{}_I&=-\frac{1}{m}\braket{k^{'2}k_I}.
\end{align}
It can be verified that the condition
\begin{align}
 \label{eq:little-def}
 \varepsilon^{KL} R^I{}_K R^J{}_L&=\varepsilon^{IJ}
\end{align}
is satisfied due to the Dirac equation~\eqref{eq:norm-spin} and the
normalization condition~\eqref{eq:dirac-little}. Note that both the primed and
un-primed spinors correspond to the same four-momentum and satisfy the
corresponding Dirac equation. Making the relation to the spin-axis notation
explicit, the components of $R$ are given by
\begin{equation}
 \label{eq:little-trafo}
\begin{aligned}
 R^1{}_1&= \frac{\braket{k^{'\flat} q}}{\braket{k^\flat q}}
 =\frac{\sbra{q'} \fmslash k\ket{ q}}{\sbraket{q' k^{'\flat}} \braket{k^\flat q}}\equiv R_{--} ,&
 R^1{}_2& = -\frac{m \sbraket{q' q }}{\sbraket{q'
      k^{'\flat}}\sbraket{k^\flat q}}\equiv R_{+-},\\
 R^2{}_1&= - \frac{m \braket{q'
     q}}{\braket{q'k^{'\flat}}\braket{k^\flat q}}\equiv R_{-+},&
R^2{}_2& 
=  \frac{\sbraket{k^{'\flat}q}}{\sbraket{k^\flat q}}
=  \frac{\bra{q'}\fmslash k\sket{q}}{\braket{q'k^{'\flat}}\sbraket{k^\flat q}}\equiv R_{++},
\end{aligned}
\end{equation}
where $k^{'\flat}$ refers to the light-cone projection of $k$ with
respect to the reference vector $q'$.

The transformation of Dirac spinors under little group
transformations~\eqref{eq:little-dirac-cov} implies the transformation of the
spinors in the spin basis~\eqref{eq:little-dirac-spin} with the matrix
\begin{align}
 \label{eq:little-dirac}
\mathcal{R}^{(\frac{1}{2})}&=
 \begin{pmatrix}
  R_{++} &R_{-+}\\ R_{+-} & R_{--}
 \end{pmatrix}, &
 \mathcal{R}^{(\frac{1}{2})-1}  
&=\begin{pmatrix}
  R_{--} &-R_{+-}\\ -R_{-+} & R_{++}
 \end{pmatrix}.
\end{align}
These results agree with those in~\cite{Schwinn:2007ee} with the
identifications $R_{++}= c_{11}$, $R_{--}= c_{22}$, $R_{-+}=- c_{12}$, and
$R_{+-}=- c_{21}$.

The matrix representation of the little-group transformation of the
polarization vectors~\eqref{eq:little-vector-spin} is found to be
\begin{equation}
 \label{eq:little-vector}
 \mathcal{R}^{(1)} =
  \begin{pmatrix}
   R_{++}^2 & -\sqrt{2} R_{-+}R_{++} & -R_{-+}^2\\
   \sqrt 2 R_{++}R_{+-} & R_{++}R_{--}+R_{-+}R_{+-} &  \sqrt
   2R_{--}R_{-+}\\
  - R_{+-}^2 & \sqrt{2} R_{+-}R_{--} & R_{--}^2
  \end{pmatrix}.
\end{equation}
It satisfies $\det{\mathcal{R}}=1$ and respects the orthonormalization
conditions of the polarization vectors,
\begin{equation}
 \mathcal{R}^{(1)}\gamma\mathcal{R}^{(1)T}=\gamma
\qquad \text{with}\qquad
 \gamma_{ss'}= \epsilon(s)\cdot\epsilon(s')
 = -\delta_{s,-s'}.
\end{equation}

\section{$Q_{\mathcal{F}}^2\neq 0$ shifts}
\label{app:q2neq0}

In an attempt to improve the large-$z$ behaviour, one may consider giving up
the requirement of light-like shifts for internal lines~\eqref{eq:shift-light}
so that propagator denominators scale like $z^2$ instead of $z^1$. The form
of the recursion relation valid for this case can be found
in~\cite{Cheung:2015cba}. Since the shift vectors $\delta k_i$ are always
light-like, momentum conservation implies that internal lines are necessarily
light-like for two-line or three-line shifts. Shifts with
$Q_{\mathcal{F}}^2\neq 0$ are thus obtained for at least four shifted lines
and are only constrained by the condition of momentum
conservation~\eqref{eq:shift-cons-fix}. Note that all shift spinors $\eta_i$
should be chosen differently since otherwise the internal shifts
$Q_{\mathcal{F},\alpha\dot\alpha}$ factorize into two Weyl spinors for some
factorization channels and the corresponding propagator denominators
``accidentally" scale like $z$. However, keeping all $\eta_i$ different
implies that spinor products of shifted spinors can scale like $z^2$ so that
the advantage of an improved large-$z$ scaling of the propagators is partially
compensated by a worse behaviour of the numerator.

A bound for the large-$z$ behaviour for generic $Q_{\mathcal{F}}^2\neq 0$
shifts can be obtained using the reasoning of
Section~\ref{sec:bound-generic}. Despite choosing all $\eta_i$
differently, contributions with $Q_{\mathcal{F}}^2=0$ can appear due to
background insertions into external shifted legs. The large-$z$ scaling of the
propagator denominators $\gamma_D$ is therefore
constrained by the mass-dimension of the denominators according
to~\cite{Cheung:2015cba}
\begin{equation}
\frac{1}{2}[D_{h,b}]\leq  \gamma_D\leq [D_{h,b}]
\end{equation}
instead of~\eqref{eq:gamma-d}. The value of $ [D_{h,b}]-\gamma_D$ is given by the
number of propagators with $Q_{\mathcal{F}}^2= 0$. A conservative upper bound
is obtained by assuming all background insertions couple through cubic
vertices and lead to a propagator with light-like shift, so that
\begin{equation}
  [D_{h,b}]-\gamma_D\leq b.
 \end{equation}
 Instead of~\eqref{eq:bound-generic}, the bound on
 the large-$z$ scaling of the amplitude becomes
\begin{align}
 \gamma&\leq 4-(h+ b)- [g]-\frac{b_\psi}{2}+[D_{h,b}]-\gamma_D- \sum_{\mathcal{H}} s_i+\sum_{\mathcal{A}} s_j\nonumber\\
& \leq 4-h -\text{min}[g]- \sum_{\mathcal{H}} s_i+\sum_{\mathcal{A}} s_j.
\end{align}
Therefore, all amplitudes in renormalizable theories are constructible from
five-line shifts with $Q_{\mathcal{F}}^2\neq 0$~\cite{Cheung:2015cba}. Since
this offers no advantage over Risager-type shifts but leads to a more
complicated recursion relation, we do not consider such
shifts any further.

\providecommand{\href}[2]{#2}\begingroup\raggedright

\endgroup


\end{document}